\documentclass[a4paper,11pt]{article}
\usepackage{jheppub} 
\usepackage{amsmath,amsfonts,amssymb,mathtools}
\usepackage{graphicx,xspace,placeins,rotating,physics,xparse}
\usepackage{lineno}
\usepackage{slashed}
\usepackage{bm}
\usepackage{booktabs}
\usepackage{comment}
\usepackage{tikz}
\usetikzlibrary{positioning,decorations.pathreplacing}
\usepackage[compat=1.1.0]{tikz-feynman}
\usepackage{subfigure}
\usepackage{xcolor,colortbl}
\usepackage{multirow,array,makecell,adjustbox}
\usepackage[inline]{enumitem}
\usepackage{dsfont}
\newcolumntype{x}[1]{>{\centering\arraybackslash\hspace{0pt}}m{#1}}

\allowdisplaybreaks
\DeclareMathAlphabet{\mathpzc}{OT1}{pzc}{m}{it}

\definecolor{nicered}{rgb}{0.7,0.1,0.1}
\definecolor{nicegreen}{rgb}{0.1,0.5,0.1}
\definecolor{niceblue}{rgb}{0.1,0.1,0.5}
\usepackage{hyperref}
\hypersetup{colorlinks,citecolor= nicegreen,linkcolor= niceblue,urlcolor= niceblue,pdfencoding=auto,psdextra}
\usepackage{breakurl}
\usepackage[capitalise]{cleveref}

\newcommand{\hhchpt}{\text{HH$\chi$PT}\xspace}

\newcommand{\nn}{\nonumber}

\newcommand{\Rv}{\mathcal{R}_v}
\newcommand{\Rvb}{\mathcal{\overline{R}}_v}
\newcommand{\Fv}{\mathcal{F}_v}
\newcommand{\Fvb}{\mathcal{\overline{F}}_v}
\newcommand{\Tv}{\mathcal{T}_v}
\newcommand{\Tvb}{\mathcal{\overline{T}}_v}
\newcommand{\Hv}{\mathcal{H}_v}
\newcommand{\Hvb}{\mathcal{\overline{H}}_v}
\newcommand{\Kv}{\mathcal{K}_v}
\newcommand{\Kvb}{\mathcal{\overline{K}}_v}

\newcommand{\Rvpb}{\mathcal{\overline{R}}_{v'}}

\newcommand{\Fvpb}{\mathcal{\overline{F}}_{v'}}

\newcommand{\Tvpb}{\mathcal{\overline{T}}_{v'}}

\newcommand{\Hvpb}{\mathcal{\overline{H}}_{v'}}

\newcommand{\Kvpb}{\mathcal{\overline{K}}_{v'}}

\newcommand{\Vpi}{\mathbb{V}}
\newcommand{\Mpi}{{\mathbb M^2}}

\newcommand{\iu}{{\rm i} }

\renewcommand{\Tr}[1]{\operatorname{Tr}\qty[#1]}
\newcommand{\plushc}{ + {\rm h.c.}\,}

\newcommand{\ffx}[2]{x^{(#1)}_{#2}}

\newcommand{\meHbHc}[1]{\matrixelement{D^{(*)}(v')}{#1}{\Bar{B}(v)}}
\newcommand{\meHbKc}[1]{\matrixelement{D^{1/2^+}(v')}{#1}{\Bar{B}(v)}}
\newcommand{\meHbFc}[1]{\matrixelement{D^{3/2^+}(v')}{#1}{\Bar{B}(v)}}
\newcommand{\meTbTc}[1]{\matrixelement{\Lambda_c(v')}{#1}{\Lambda_b(v)}}
\newcommand{\meTbRc}[1]{\matrixelement{\Lambda_{c1}^{(*)}(v')}{#1}{\Lambda_b(v)}}

\preprint{CALT-TH-2025-038}

\title{\boldmath Radiative Semileptonic Decays of Beautiful Hadrons}

\author{Federico Cima,}
\emailAdd{fcima@caltech.edu}
\author{Michele Papucci}
\emailAdd{mpapucci@caltech.edu}
\affiliation{Walter Burke Institute for Theoretical Physics,\\ 
 California Institute of Technology, Pasadena, CA 91125, USA}

\abstract{
We derive predictions for the hadronic matrix elements of radiative semileptonic decays of beautiful hadrons within Heavy Quark Effective Theory (HQET), relevant for future measurements at Belle II and LHCb. Our study considers $\Lambda_b\to \Lambda_c, \Lambda_{c1}^{(*)}$ and $\bar{B} \to D^{(*)}, D^{**}$ transitions. 
The symmetries of HQET highly constrain the number of structure-dependent form factors in all cases. 
In the soft and sub-leading soft regions, all the form factors are fully determined in terms of non-radiative Isgur-Wise functions and the magnetic dipole moments of the heavy hadrons. The structure of higher order corrections is also briefly discussed.}

\begin{document}
\maketitle
\flushbottom
\section{Introduction}
In the limit where the bottom and charm quark masses are much heavier than the scale of non-perturbative strong dynamics, i.e $m_{b,c}\gg\Lambda_{QCD}$, QCD exhibits a new symmetry: heavy quark spin-flavor symmetry (HQS)~\cite{Nussinov:1986hw, Isgur:1989vq, Isgur:1990yhj}. Such symmetry acts on heavy quarks with the same four-velocity, and it is the founding principle of the \emph{Heavy Quark Effective Theory} (HQET).
This is among the most powerful tools to unveil non-perturbative properties of heavy hadrons, and it has been famously used to constrain the number of form factors needed to parameterize hadronic matrix elements involving heavy quarks. 

In this paper, we study radiative semileptonic decays of bottom hadrons, as initiated in~\cite{Papucci:2021ztr}. We focus on $\Lambda_b$ decays to $\Lambda_c$ and $\Lambda_{c1}^{(*)}$ baryons and on the $\bar{B}$ decays to $D^{**}$ mesons, extending the results of~\cite{Papucci:2021ztr} for the case of $\bar{B} \rightarrow D^{(*)}$ decays.
The presence of an additional photon allows for testing the predictability of HQET in more complicated kinematic settings than usual semileptonic decays, without involving additional hadrons and the associated calculational uncertainties. Introducing a new four-momentum, namely the photon four-momentum $k$, new kinematical invariants besides $w\equiv v\cdot v'$ can be constructed, i.e. $v\cdot k$ and $v'\cdot k$. Therefore, form factors for these processes will generally depend on three Lorentz scalars rather than one. 

A second motivation is that Belle II aims to enable precision studies of radiative semileptonic $\bar{B}$ decays, reaching a level comparable to that achieved for semileptonic $\bar{B}$ decays by earlier B factories, such as Belle, BaBar, and LHCb. To do so, a solid theoretical understanding and parameterization of these corrections is required. Additionally, any radiative semileptonic decay represents an irreducible background for non-radiative semileptonic measurements in kinematic regions where the photon remains undetected.  This is particularly relevant in the case of LHCb, where the hadronic environment leads to higher energy thresholds for photon detection. Thus, our study serves as groundwork for modeling such backgrounds and consequently for a precise interpretation of future experimental results.

The full HQET prediction for the hadronic matrix elements of interest includes a number of $k$-dependent operators describing at the non-perturbative level the time-ordered product between the photon field and the hadronic current. This properly characterizes the radiative semileptonic decays even away from the purely soft limit, as studied previously in the literature for the $\bar{B}\to D^{(*)}$ transition~\cite{Becirevic:2009fy, Bernlochner:2010yd, deBoer:2018ipi}, in terms of \emph{structure-dependent} (SD) functions. The parameterization remains valid in the regime of validity HQS, namely $v^{(\prime)}\cdot k \ll 2 m_{b,c}$. 
HQS highly constrains the number of independent SD functions. In~\cite{Papucci:2021ztr} it was found that for $\bar{B}\rightarrow D^{(*)}$ at leading order (LO) in $1/m_{b,c}$ only four SD functions were necessary instead of 32 for the general Lorentz-invariant case.

Further insight on the structure of such form factors can be assessed by matching the HQET matrix elements onto those computed in \emph{Heavy Hadron Chiral Perturbation Theory} (\hhchpt), by focusing on the soft and sub-leading soft regions.
This matching procedure leads to a further reduction in the set of new structure constants needed to describe these processes in these kinematic regimes, which are also the interesting ones for experiments. The inclusion of the sub-leading soft region extends what has been currently included in MonteCarlo event generators used in experimental analyses, such as \texttt{PHOTOS}~\cite{Barberio:1993qi}, which is inherently limited to the point-particle approximation.
Remarkably, we find that in this procedure, HQS, particularly its invariance under the reparameterization (RPI) of the residual momenta, is incredibly powerful in constraining the form of the SD functions.

The outline of the paper is the following. In \cref{sec:HQET parameterization} we work out the HQET parameterization of the amplitude for radiative semileptonic decays for phenomenologically interesting B-baryon and B-meson decay channels. Then, \cref{sec:soft limit} presents the matching between HQET and \hhchpt at leading and sub-leading soft orders, providing reduced sets of structure-dependent form factors that are relevant in these regimes. We give our final remarks and conclude in \cref{sec:Summary}.

\section{HQET parameterization}
\label{sec:HQET parameterization}
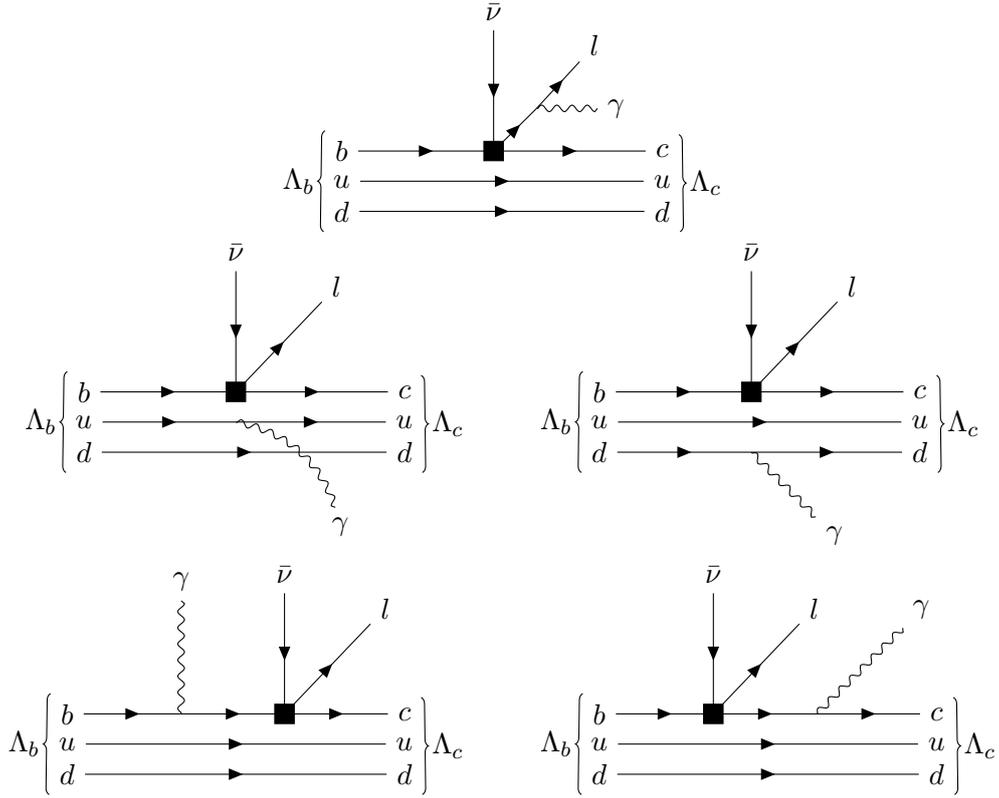
\begin{figure}
    \centering
        \begin{tikzpicture}[baseline=(a.base)]
        \begin{feynman}[inline=(a.base)]
        \vertex (a) {\(b\)};
        \vertex [right= 2cm of a] (b);
        \node at (b) [shape=rectangle, draw=black, fill, minimum size=2pt];
        \vertex [right= 2cm of b] (c) {\(c\)};
        \vertex [below= 0.4cm of a] (d) {\(u\)};
        \vertex [below= 0.4cm of c] (e) {\(u\)};
        \vertex [below= 0.4cm of d] (f) {\(d\)};
        \vertex [below= 0.4cm of e] (g) {\(d\)};
        \vertex [above right= 0.8cm of b] (h);
        \vertex [above right= 0.8cm of h] (i) {\(l\)};
        \vertex [above = 1.6cm of b] (l) {\(\bar{\nu}\)};
        \vertex [right= 0.8cm of h] (m) {\(\gamma\)};
        \diagram* {
        (a) -- [fermion] (b) -- [fermion] (c);
        (d) -- [fermion] (e);
        (f) -- [fermion] (g);
        (b) -- [fermion] (h) -- [fermion] (i);
        (l) -- [fermion] (b);
        (h) -- [photon] (m);
        };
        \draw [decoration={brace}, decorate] (f.south west) -- (a.north west) node [pos=0.5, left] {\(\Lambda_b\)};
        \draw [decoration={brace}, decorate] (c.north east) -- (g.south east) node [pos=0.5, right] {\(\Lambda_c\)};
        \end{feynman}
        \end{tikzpicture}\\
        \begin{tikzpicture}[baseline=(a.base)]
        \begin{feynman}[inline=(a.base)]
        \vertex (a) {\(b\)};
        \vertex [right= 2cm of a] (b);
        \node at (b) [draw, fill];
        \vertex [right= 2cm of b] (c) {\(c\)};
        \vertex [below= 0.4cm of a] (d) {\(u\)};
        \vertex [below= 0.4cm of c] (e) {\(u\)};
        \vertex [below= 0.4cm of d] (f) {\(d\)};
        \vertex [below= 0.4cm of e] (g) {\(d\)};
        \vertex [above right= 1.6cm of b] (h) {\(l\)};
        \vertex [below = 0.4cm of b] (i);
        \vertex [above = 1.6cm of b] (l) {\(\bar{\nu}\)};
        \vertex [below right= 1.6cm of i] (m) {\(\gamma\)};
        \diagram* {
        (a) -- [fermion] (b) -- [fermion] (c);
        (d) -- [fermion] (i) -- [fermion] (e);
        (f) -- [fermion] (g);
        (b) -- [fermion] (h);
        (l) -- [fermion] (b);
        (i) -- [photon, bend left] (m);
        };
        \draw [decoration={brace}, decorate] (f.south west) -- (a.north west) node [pos=0.5, left] {\(\Lambda_b\)};
        \draw [decoration={brace}, decorate] (c.north east) -- (g.south east) node [pos=0.5, right] {\(\Lambda_c\)};
        \end{feynman}
        \end{tikzpicture}
        \hspace{0.5cm}
        \begin{tikzpicture}[baseline=(a.base)]
        \begin{feynman}[inline=(a.base)]
        \vertex (a) {\(b\)};
        \vertex [right= 2cm of a] (b);
        \node at (b) [draw, fill];
        \vertex [right= 2cm of b] (c) {\(c\)};
        \vertex [below= 0.4cm of a] (d) {\(u\)};
        \vertex [below= 0.4cm of c] (e) {\(u\)};
        \vertex [below= 0.4cm of d] (f) {\(d\)};
        \vertex [below= 0.4cm of e] (g) {\(d\)};
        \vertex [above right= 1.6cm of b] (h) {\(l\)};
        \vertex [below = 0.8cm of b] (i);
        \vertex [above = 1.6cm of b] (l) {\(\bar{\nu}\)};
        \vertex [below right= 1.2cm of i] (m) {\(\gamma\)};
        \diagram* {
        (a) -- [fermion] (b) -- [fermion] (c);
        (d) -- [fermion] (e);
        (f) -- [fermion] (i) -- [fermion] (g);
        (b) -- [fermion] (h);
        (l) -- [fermion] (b);
        (i) -- [photon] (m);
        };
        \draw [decoration={brace}, decorate] (f.south west) -- (a.north west) node [pos=0.5, left] {\(\Lambda_b\)};
        \draw [decoration={brace}, decorate] (c.north east) -- (g.south east) node [pos=0.5, right] {\(\Lambda_c\)};
        \end{feynman}
        \end{tikzpicture}
        \hspace{0.5cm}
        \begin{tikzpicture}[baseline=(a.base)]
        \begin{feynman}[inline=(a.base)]
        \vertex (a) {\(b\)};
        \vertex [right= 1.5cm of a] (i);
        \vertex [right= 1.36cm of i] (b);
        \node at (b) [draw, fill];
        \vertex [right = 1.36cm of b] (c) {\(c\)};
        \vertex [below= 0.4cm of a] (d) {\(u\)};
        \vertex [below= 0.4cm of c] (e) {\(u\)};
        \vertex [below= 0.4cm of d] (f) {\(d\)};
        \vertex [below= 0.4cm of e] (g) {\(d\)};
        \vertex [above right= 1.6cm of b] (h) {\(l\)};
        \vertex [above = 1.6cm of b] (l) {\(\bar{\nu}\)};
        \vertex [above = 1.5cm of i] (m) {\(\gamma\)};
        \diagram* {
        (a) -- [fermion] (i) -- [fermion] (b) -- [fermion] (c);
        (d) -- [fermion] (e);
        (f) -- [fermion] (g);
        (b) -- [fermion] (h);
        (l) -- [fermion] (b);
        (i) -- [photon] (m);
        };
        \draw [decoration={brace}, decorate] (f.south west) -- (a.north west) node [pos=0.5, left] {\(\Lambda_b\)};
        \draw [decoration={brace}, decorate] (c.north east) -- (g.south east) node [pos=0.5, right] {\(\Lambda_{c}\)};
        \end{feynman}
        \end{tikzpicture}
        \hspace{0.5cm}
        \begin{tikzpicture}[baseline=(a.base)]
        \begin{feynman}[inline=(a.base)]
        \vertex (a) {\(b\)};
        \vertex [right= 1.5cm of a] (b);
        \node at (b) [draw, fill];
        \vertex [right= 1.36cm of b] (i);
        \vertex [right = 1.36cm of i] (c) {\(c\)};
        \vertex [below= 0.4cm of a] (d) {\(u\)};
        \vertex [below= 0.4cm of c] (e) {\(u\)};
        \vertex [below= 0.4cm of d] (f) {\(d\)};
        \vertex [below= 0.4cm of e] (g) {\(d\)};
        \vertex [above right= 1.6cm of b] (h) {\(l\)};
        \vertex [above = 1.6cm of b] (l) {\(\bar{\nu}\)};
        \vertex [above right= 1.6cm of i] (m) {\(\gamma\)};
        \diagram* {
        (a)  -- [fermion] (b) -- [fermion] (i) -- [fermion] (c);
        (d) -- [fermion] (e);
        (f) -- [fermion] (g);
        (b) -- [fermion] (h);
        (l) -- [fermion] (b);
        (i) -- [photon] (m);
        };
        \draw [decoration={brace}, decorate] (f.south west) -- (a.north west) node [pos=0.5, left] {\(\Lambda_b\)};
        \draw [decoration={brace}, decorate] (c.north east) -- (g.south east) node [pos=0.5, right] {\(\Lambda_{c}\)};
        \end{feynman}
        \end{tikzpicture}
        \caption{Feynman diagrams contributing to the decay process $\Lambda_b\to \Lambda_c l\bar{\nu}\gamma$. The same diagrams are also the ones relevant for the decay process $\Lambda_b\to \Lambda_{c1}^{(*)} l\bar{\nu}\gamma$. The black square indicates a weak current insertion.}
        \label{fig:baryon Feynman diagrams}
\end{figure}

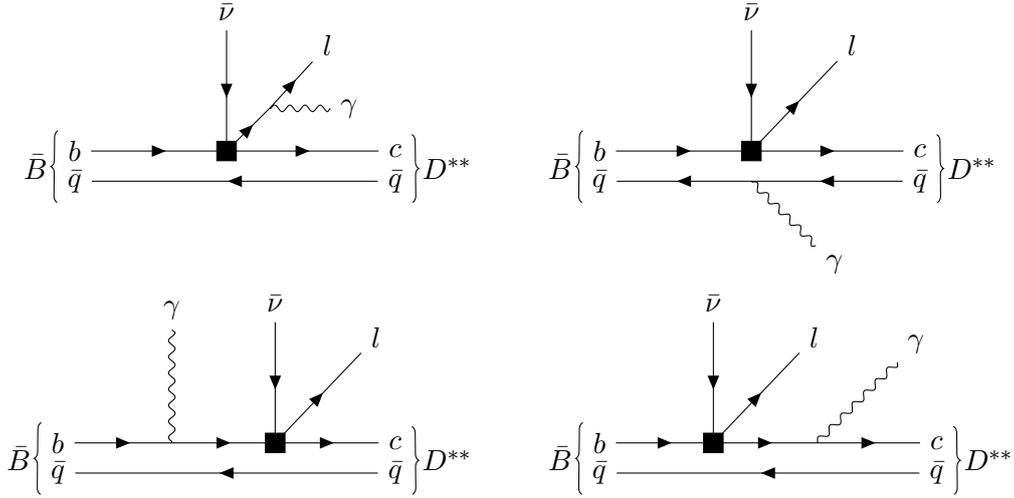
\begin{figure}
    \centering
        \begin{tikzpicture}[baseline=(a.base)]
        \begin{feynman}[inline=(a.base)]
        \vertex (a) {\(b\)};
        \vertex [right= 2cm of a] (b);
        \node at (b) [draw, fill];
        \vertex [right= 2cm of b] (c) {\(c\)};
        \vertex [below= 0.4cm of a] (d) {\(\bar{q}\)};
        \vertex [below= 0.4cm of c] (e) {\(\bar{q}\)};
        \vertex [above right= 0.8cm of b] (h);
        \vertex [above right= 0.8cm of h] (i) {\(l\)};
        \vertex [above = 1.6cm of b] (l) {\(\bar{\nu}\)};
        \vertex [right= 0.8cm of h] (m) {\(\gamma\)};
        \diagram* {
        (a) -- [fermion] (b) -- [fermion] (c);
        (e) -- [fermion] (d);
        (b) -- [fermion] (h) -- [fermion] (i);
        (l) -- [fermion] (b);
        (h) -- [photon] (m);
        };
        \draw [decoration={brace}, decorate] (d.south west) -- (a.north west) node [pos=0.5, left] {\(\bar{B}\)};
        \draw [decoration={brace}, decorate] (c.north east) -- (e.south east) node [pos=0.5, right] {\(D^{**}\)};
        \end{feynman}
        \end{tikzpicture}
        \hspace{0.5cm}
        \begin{tikzpicture}[baseline=(a.base)]
        \begin{feynman}[inline=(a.base)]
        \vertex (a) {\(b\)};
        \vertex [right= 2cm of a] (b);
        \node at (b) [draw, fill];
        \vertex [right= 2cm of b] (c) {\(c\)};
        \vertex [below= 0.4cm of a] (d) {\(\bar{q}\)};
        \vertex [below= 0.4cm of c] (e) {\(\bar{q}\)};
        \vertex [above right= 1.6cm of b] (h) {\(l\)};
        \vertex [below = 0.4cm of b] (i);
        \vertex [above = 1.6cm of b] (l) {\(\bar{\nu}\)};
        \vertex [below right= 1.2cm of i] (m) {\(\gamma\)};
        \diagram* {
        (a) -- [fermion] (b) -- [fermion] (c);
        (e) -- [fermion] (i) -- [fermion] (d);
        (b) -- [fermion] (h);
        (l) -- [fermion] (b);
        (i) -- [photon] (m);
        };
        \draw [decoration={brace}, decorate] (d.south west) -- (a.north west) node [pos=0.5, left] {\(\bar{B}\)};
        \draw [decoration={brace}, decorate] (c.north east) -- (e.south east) node [pos=0.5, right] {\(D^{**}\)};
        \end{feynman}
        \end{tikzpicture}\\
        \begin{tikzpicture}[baseline=(a.base)]
        \begin{feynman}[inline=(a.base)]
        \vertex (a) {\(b\)};
        \vertex [right= 1.5cm of a] (i);
        \vertex [right= 1.36cm of i] (b);
        \node at (b) [draw, fill];
        \vertex [right = 1.36cm of b] (c) {\(c\)};
        \vertex [below= 0.4cm of a] (d) {\(\bar{q}\)};
        \vertex [below= 0.4cm of c] (e) {\(\bar{q}\)};
        \vertex [above right= 1.6cm of b] (h) {\(l\)};
        \vertex [above = 1.6cm of b] (l) {\(\bar{\nu}\)};
        \vertex [above = 1.5cm of i] (m) {\(\gamma\)};
        \diagram* {
        (a) -- [fermion] (i) -- [fermion] (b) -- [fermion] (c);
        (e) -- [fermion] (d);
        (b) -- [fermion] (h);
        (l) -- [fermion] (b);
        (i) -- [photon] (m);
        };
        \draw [decoration={brace}, decorate] (d.south west) -- (a.north west) node [pos=0.5, left] {\(\bar{B}\)};
        \draw [decoration={brace}, decorate] (c.north east) -- (e.south east) node [pos=0.5, right] {\(D^{**}\)};
        \end{feynman}
        \end{tikzpicture}
        \hspace{0.5cm}
        \begin{tikzpicture}[baseline=(a.base)]
        \begin{feynman}[inline=(a.base)]
        \vertex (a) {\(b\)};
        \vertex [right= 1.5cm of a] (b);
        \node at (b) [draw, fill];
        \vertex [right= 1.36cm of b] (i);
        \vertex [right = 1.36cm of i] (c) {\(c\)};
        \vertex [below= 0.4cm of a] (d) {\(\bar{q}\)};
        \vertex [below= 0.4cm of c] (e) {\(\bar{q}\)};
        \vertex [above right= 1.6cm of b] (h) {\(l\)};
        \vertex [above = 1.6cm of b] (l) {\(\bar{\nu}\)};
        \vertex [above right= 1.5cm of i] (m) {\(\gamma\)};
        \diagram* {
        (a)  -- [fermion] (b) -- [fermion] (i) -- [fermion] (c);
        (e) -- [fermion] (d);
        (b) -- [fermion] (h);
        (l) -- [fermion] (b);
        (i) -- [photon] (m);
        };
        \draw [decoration={brace}, decorate] (d.south west) -- (a.north west) node [pos=0.5, left] {\(\bar{B}\)};
        \draw [decoration={brace}, decorate] (c.north east) -- (e.south east) node [pos=0.5, right] {\(D^{**}\)};
        \end{feynman}
        \end{tikzpicture}
        \caption{Feynman diagrams contributing to the decay process $\bar{B}\to D^{**} l\bar{\nu}\gamma$. The black square indicates the weak current insertion.}
        \label{fig:meson Feynman diagrams}
\end{figure}

In this section, we derive the HQET parameterization of the amplitudes for all the decay channels we consider. We first look at the baryon case in \cref{subsec:Baryons} and then to the meson case in \cref{subsec:Mesons}. 
The matrix element $\mathcal{M}$ can be decomposed into three different contributions depending on where the photon is attached, namely the photon emission from the charged lepton, the heavy quarks, or the light QCD degrees of freedom (the brown muck).
Representative Feynman diagrams for each of these processes are shown in \cref{fig:baryon Feynman diagrams,fig:meson Feynman diagrams}.
Since all the amplitudes have a photon polarization appended, we factor it out by defining $\mathcal{M}\equiv \epsilon_\mu \mathcal{M}^\mu$ in the following. 
At LO in the HQET mass expansion, the contributions to $\mathcal{M}^\mu$ from the heavy quarks and brown muck emissions will be given by the matrix elements of the time-ordered product of the LO electromagnetic currents
\begin{equation}
J^\mu_b = v^\mu e_b \bar b_v b_v, \ \ \ J^\mu_c = v^{\prime \mu} e_c \bar c_{v'} c_{v'}, \ \ \ J^\mu_{light} = \sum_{i = u,d,s} e_i \bar q_i \gamma^\mu q_i,
\end{equation}
with the semileptonic weak operator
\begin{equation}
\mathcal{O}_W =\frac{G_F V_{cb}}{\sqrt 2} \alpha_{\Gamma \Gamma'}(\bar c_{v'} \Gamma b_v)(\bar u_\nu \Gamma^\prime v_\ell),
\end{equation}
where we have kept the Dirac structure generic to include also possible new physics effects, since the parameterization of the electromagnetic form factors is independent of the Dirac structure of the weak current operator.  $\Gamma^{(\prime)}=S, P, V, A, T=1, \gamma^5, \gamma^\nu,\gamma^\nu\gamma^5, \sigma^{\mu\nu}$. For the specific case of the Standard Model (SM), we have $\Gamma=\Gamma'=V-A$, and $\alpha_{\Gamma \Gamma'}=1$.

If the photon is emitted from the charged lepton, then the matrix element can generically be written as 
\begin{equation}
    \frac{\mathcal{M}_{lepton}^\mu}{\sqrt{m_{H_b}m_{H_c}}}=e_l\frac{G_F e V_{cb}\alpha_{\Gamma \Gamma'}}{\sqrt{2}}\frac{F_\Gamma}{2 p_l\cdot k}\bar{u} \gamma^\mu (\slashed{p_l}+\slashed{k}+m_l)\Gamma' v,
    \label{eq:lepton matrix element}
\end{equation}
where $m_{H_b}$ and $m_{H_c}$ are the masses of the generic bottom and charm hadron respectively, $e_l=-1$ is the electromagnetic charge of the lepton, $p_l$ is the lepton four-momentum, $k$ is the photon four-momentum and $F_\Gamma$ is the hadronic matrix element of the non-radiative decay\footnote{$F_\Gamma$ is evaluated at shifted kinematic $Q^2 \rightarrow Q^2 + 2 k \cdot Q$, but the two kinematic points differ only at higher order in the heavy quark mass expansion.}:
\begin{equation}
F_\Gamma = \mel{H^{v'}_c}{\bar c_{v'} \Gamma b_v}{H^v_b}.
\end{equation}

The second class of diagrams comprises instead all the possible photon radiations from the brown muck and its corresponding amplitude is
\begin{equation}
     \frac{\mathcal{M}_{light}^\mu}{\sqrt{m_{H_b}m_{H_c}}}=\frac{G_F e V_{cb}}{\sqrt{2}}M_\Gamma^{\mu}\bar{u}\Gamma'v,
     \label{eq:light matrix element}
\end{equation}
where $M_\Gamma^{\mu}$ is given by 
\begin{equation}
M^\mu_\Gamma (k)=  \mel{H^{v'}_c}{\bar c_{v'} \Gamma b_v \circ_k J_{light}^\mu}{H^v_b}.
\end{equation}
where we have denoted with $\circ_k$ the Fourier transform of the time-ordered product at finite momentum $k$:
\begin{equation}
O \circ_k O' \equiv \int \frac{\dd[4]{x}}{(2\pi)^4} e^{i k \cdot x} {\rm T}\qty{ O(0) O'(x)}
\end{equation}

The third and final class of diagrams are those containing the two possible photons radiations from the heavy quarks. The contribution to the total amplitude from these diagrams reads
\begin{equation}
     \frac{\mathcal{M}_{heavy}^\mu}{\sqrt{m_{H_b}m_{H_c}}}=\frac{G_F e V_{cb}}{\sqrt{2}}\qty[e_b v^\mu G_{\Gamma, b}+e_c v'^\mu G_{\Gamma, c}]\bar{u}\Gamma' v,
     \label{eq:heavy matrix element}
\end{equation}
where $e_b$ and $e_c$ are the electromagnetic charges of the bottom and charm quarks, while $G_{\Gamma,b}$ and $G_{\Gamma,c}$ are hadronic matrix elements describing the emission from the charm and bottom quarks and are defined as:
\begin{align}
e_b v^\mu G_{\Gamma,b} (k) &= \mel{H^{v'}_c}{\bar c_{v'} \Gamma b_v \circ_k J_{b}^\mu}{H^v_b}, \nn\\
e_c v^{\prime\mu} G_{\Gamma,c} (k) &= \mel{H^{v'}_c}{J_{c}^\mu \circ_k \bar c_{v'} \Gamma b_v}{H^v_b}.
\end{align}

Following~\cite{Papucci:2021ztr}, the total amplitude can be written in a compact form as
\begin{equation}
    \frac{\mathcal{M}^\mu}{\sqrt{m_{H_b}m_{H_c}}}=\frac{G_F e V_{cb}}{\sqrt{2}}\bigg[K_\Gamma^{\mu}\bar{u}\Gamma' v
    +(e_b-e_c)\frac{F_\Gamma}{2 p_l\cdot k}\bar{u} \gamma^\mu (\slashed{p_l}+\slashed{k}+m_l)\Gamma' v\bigg],
   \label{eq:total amplitude}
\end{equation}
where we defined the tensor $K_\Gamma^{\mu}\equiv M_\Gamma^{\mu}+e_b v^\mu G_{\Gamma, b}+e_c v'^\mu G_{\Gamma, c}$ and we have written $e_l=e_b-e_c$.  
$\mathcal{M}^\mu$ in \cref{eq:total amplitude} must satisfy the \emph{Ward identity} (WI), $k_\mu \mathcal{M}^\mu=0$, leading to constraints on the structure of the hadronic matrix elements. As noted in~\cite{Papucci:2021ztr}, since only $M^{\mu}_\Gamma$ can depend on the charge of the light quarks, and $e_{b,c}$ can be varied independently, one can impose the WI separately on $M^{\mu}_\Gamma$ and the terms proportional to $e_{b,c}$, obtaining the three equations:
\begin{subequations}
    \begin{align}
        k_\mu M_\Gamma^{\mu}&=0,\\*
        (k\cdot v) G_{\Gamma, b} + F_\Gamma &=0,\\*
        (k\cdot v') G_{\Gamma, c} - F_\Gamma &=0.
    \end{align}
    \label{eq:Ward identity}
\end{subequations}

The explicit parameterizations of \cref{eq:total amplitude} for the different decay channels will be presented in the next two subsections. To avoid confusion, we will denote all objects (i.e. amplitudes, hadronic matrix elements and form factors) across different transitions with the same name, but differentiate specific transitions by using the following decorations:
\begin{equation}
    \Lambda_b\to \Lambda_c: \mathcal{O},\quad \Lambda_b\to \Lambda_{c1}^{(*)}: \breve{\mathcal{O}},\quad
    \bar{B}\to D^{(*)}: \bar{\mathcal{O}}, \quad\bar{B}\to D^{1/2^+}: \tilde{\mathcal{O}}, \quad\bar{B}\to D^{3/2^+}: \hat{\mathcal{O}}.
\end{equation}

\subsection{Baryons}
\label{subsec:Baryons}
In this subsection, we start by exploring the radiative semileptonic decays of the $\Lambda_b$ baryon to $\Lambda_c$ and $\Lambda_{c1}^{(*)}$ baryons. The procedure will closely follow the one employed in~\cite{Papucci:2021ztr}, where the study of radiative semileptonic $B$-hadron decays has been recently initiated considering $\bar{B}\to D^{(*)}l\bar{\nu}\gamma$. 

\subsubsection{$\Lambda_b\to \Lambda_c l \Bar{\nu}\gamma$  form factors}
 As for the $\Bar{B}\to D^{(*)}l\Bar{\nu}$ case~\cite{Manohar:2000dt}, in the HQ limit, $F_\Gamma$ depends on a single Isgur-Wise (IW) function $\zeta(w)$,
\begin{equation}
    F_\Gamma=\zeta(w)\meTbTc{\overline{\mathcal{T}}_{v'}^{(c)}\Gamma \mathcal{T}_v^{(b)}}=\zeta (w) \overline{T}_{v'}^{(c)}\Pi'_+\Gamma \Pi_+ T_v^{(b)},
\end{equation}
where the matrix element is already written in terms of the super-field $\mathcal{T}$, while $T$ is a the Dirac spinor and $\Pi_+,\Pi'_+\equiv (1+\slashed{v})/2,(1+\slashed{v'})/2$. HQS implies the normalization condition at \emph{zero recoil}: $\zeta(1)=1$.
Moving to the photon emission from the brown muck, the most general expression for the hadronic matrix element $M_\Gamma$ consistent with HQS is 
\begin{equation}
    M_{\Gamma}^{\mu}=\mathcal{X}^\mu \meTbTc{\overline{\mathcal{T}}_{v'}^{(c)}\Gamma \mathcal{T}_v^{(b)}}=\mathcal{X}^\mu \overline{T}_{v'}^{(c)}\Pi'_+\Gamma\Pi_+ T_v^{(b)},
\end{equation}
where 
\begin{equation}
    \mathcal{X}^\mu=\ffx{1}{} v^\mu+\ffx{2}{}v'^\mu.
\end{equation}
We have dropped the term proportional to $k^\mu$, that would produce a vanishing contribution to the matrix element. Notice that, differently to the meson case, we cannot insert any Dirac structure outside the fermion current as the brown muck is in an angular momentum singlet state. The $\ffx{j}{}$ are scalar form factors depending on $w$, $v\cdot k$ and $v'\cdot k$. 
They have the same mass dimension $\qty[\ffx{1-2}{}]=-1$.

Finally, the hadronic matrix elements that characterized the photon emission from the heavy quarks $G_{\Gamma,Q}$ can be written as
\begin{equation}
    G_{\Gamma,Q}=y_Q^{(1)} \meTbTc{\overline{\mathcal{T}}_{v'}^{(c)}\Gamma \mathcal{T}_v^{(b)}}=y_Q^{(1)}\overline{T}_{v'}^{(c)}\Pi'_+\Gamma\Pi_+ T_v^{(b)},
\end{equation}
where, as the $\ffx{j}{}$, $y_Q^{(1)}$ is a scalar form factor depending on $w$, $v\cdot k$ and $v'\cdot k$ with mass dimension $\qty[y^{(1)}_Q]=-1$.  

By using HQS we are left with four new form factors (besides the Isgur-Wise function $\zeta(w)$): $x^{(1-2)}$ and $y^{(1)}_{b,c}$. This is still an overcomplete basis, as we still have to impose the WI of \cref{eq:Ward identity}, which provide three additional constraints. 
Therefore, we are left with just one additional new structure function:
\begin{subequations}
    \begin{align}
        x^{(2)}&=-\frac{v\cdot k}{v'\cdot k}x^{(1)},\\
        y_b^{(1)}&=-\frac{\zeta}{v\cdot k},\\
        y_c^{(1)}&=\frac{\zeta}{v'\cdot k}.
    \end{align}
\end{subequations}
We choose the new structure function to be
\begin{equation}
    \varrho^{(1)}\equiv (v\cdot k) x^{(1)}.
\end{equation}
with $\varrho^{(1)}$ a dimensionless function of $w$, $v\cdot k$, $v'\cdot k $.
With this choice, $\mathcal{X}^\mu$ simplifies to
\begin{equation}\label{eq:XT}
    \mathcal{X}^\mu=\varrho^{(1)}\qty(\frac{v^\mu}{v\cdot k}-\frac{v'^\mu}{v'\cdot k}),
\end{equation}
rendering manifest that $\mathcal{X}^\mu$ is proportional to a soft factor, as explicitly shown in \cref{sec:soft limit}.

\subsubsection{$\Lambda_b\to \Lambda_{c1}^{(*)} l \Bar{\nu}\gamma$ form factors}
We now discuss the radiative semileptonic $\Lambda_b$ decays to the $p$-wave excited state of the $\Lambda_c$ baryon, namely the HQS doublet $\Lambda_{c1}^{(*)}$. The non-radiative decay channelsl $\Lambda_b\to \Lambda_{c1}^{(*)}l\bar{\nu}\to \Lambda_c \pi^a \pi^b l \bar{\nu}$, first studied in~\cite{Cho:1994vg}, are potential targets for studying $b \rightarrow c$ transitions at LHCb and they  present puzzles in understanding current experimental measurements and lattice results~\cite{Bernlochner:2022hyz,cim:hig}.

We follow an analogous procedure to the previous case (upon changing the mass of the $\Lambda_c$ with the spin-averaged mass of the $\Lambda_{c1}^{(*)}$, and using the ``$\;\breve{\mbox{}}\;$'' notation for the various quantities).
$\breve{F}_\Gamma$ can still be written in terms of a single Isgur-Wise function $\sigma(w)$
\begin{equation}
    \breve{F}_{\Gamma}=\sigma(w)\meTbRc{v_\mu \overline{\mathcal{R}}_{v'}^{(c)\mu}\Gamma \mathcal{T}_v^{(b)}},
\end{equation}
where $\mathcal{R}_{v'}^{(c)\mu}$ is the super-field associated to the $\Lambda_{c1}^{(*)}$ multiplet, satisfying $v'\cdot \mathcal{R}=0$. At zero recoil, this matrix element vanishes, so that $\sigma(1)$ is not constrained to unity by HQS.
The hadronic matrix element associated with the photon radiation from the light quarks $\breve{M}_\Gamma$ is instead
\begin{equation}
    \breve{M}^{\mu}_{\Gamma}=\breve{\mathcal{X}}^\mu_{\,\sigma} \meTbRc{\overline{\mathcal{R}}_{v'}^{(c)\sigma}\Gamma \mathcal{T}_v^{(b)}},
\end{equation}
where 
\begin{equation}
    \breve{\mathcal{X}}^\mu_{\,\sigma}=\breve{x}^{(1)}\delta^\mu_\sigma+\breve{x}^{(2)}v_\sigma v^\mu+\breve{x}^{(3)}v_\sigma v'^\mu+\breve{x}^{(4)}k_\sigma v^\mu+\breve{x}^{(5)}k_\sigma v'^\mu.
\end{equation}
The $\breve{x}^{(j)}$ are scalar form factors depending on $w$, $v\cdot k$ and $v'\cdot k$. Their mass dimensions are $\qty[\breve{x}^{(1-3)}]=-1$ and $\qty[\breve{x}^{(4-5)}]=-2$. We have dropped all the terms proportional to $k^\mu$ and all the terms that violate P and CP invariance.
Finally, the hadronic matrix elements $G'_{\Gamma, Q}$ can be written as
\begin{equation}
    \breve{G}_{\Gamma,Q}=\breve{\mathcal{Y}}_{Q\mu}\meTbRc{\overline{\mathcal{R}}_{v'}^{(c)\mu}\Gamma \mathcal{T}_v^{(b)}},
\end{equation}
where
\begin{equation}
    \breve{\mathcal{Y}}_{Q \mu}=\breve{y}^{(1)}_Q v_\mu+\breve{y}^{(2)}_Q k_\mu.
\end{equation}
The $\breve{y}^{(j)}_Q$ are scalar form factors depending on $w$, $v\cdot k$ and $v'\cdot k$, with mass dimensions $\qty[\breve{y}^{(1)}_Q]=-1$ and $\qty[\breve{y}^{(2)}_Q]=-2$.

HQS constraints the number of new form factors, besides the Isgur-Wise function $\sigma(w)$, down to an overcomplete set of nine form factors: $\breve{x}^{(1-5)}$ and $\breve{y}^{(1-2)}_{b,c}$. Imposing the WI, \cref{eq:Ward identity}, we get:
\begin{subequations}
    \begin{align}
        \breve{x}^{(1)}&=-(v\cdot k)\breve{x}^{(4)}-(v'\cdot k)\breve{x}^{(5)},\\
        \breve{x}^{(3)}&=-\frac{v\cdot k}{v'\cdot k}\breve{x}^{(2)},\\
        \breve{y}^{(1)}_b&=-\frac{\sigma}{v\cdot k},\\
        \breve{y}^{(1)}_c&=\frac{\sigma}{v'\cdot k},\\
        \breve{y}^{(2)}_{b,c}&=0.
    \end{align}
\end{subequations}
Thus, there are only three new structure functions:
\begin{subequations}
    \begin{align}
        \breve{\varrho}^{(1)}&\equiv (v\cdot k) \breve{x}^{(2)},\\
        \breve{\varrho}^{(2)}&\equiv -(v\cdot k) \breve{x}^{(4)},\\
        \breve{\varrho}^{(2')}&\equiv -(v' \cdot k) \breve{x}^{(5)},
    \end{align}
\end{subequations}
all functions of $w$, $v \cdot k$ and $v' \cdot k$. $\breve{\varrho}^{(1)}$ is dimensionless, while $\breve{\varrho}^{(2,2')}$ start at mass dimension $-1$. The rationale behind this specific choice of normalizations will be clear upon studying their infrared behavior in \cref{sec:soft limit}.
In this basis, $\breve{\mathcal{X}}$ simplifies to
\begin{equation}\label{eq:XR}
    \breve{\mathcal{X}}^\mu_{\,\sigma}= \breve{\varrho}^{(1)}\qty(\frac{v^\mu}{v\cdot k}-\frac{v'^\mu}{v'\cdot k})v_\sigma+\breve{\varrho}^{(2)} \mathcal{P}^\mu_\sigma+\breve{\varrho}^{(2')} \mathcal{P}^{\prime\mu}_\sigma.
\end{equation}
where we have defined the tensor $\mathcal{P}^{(\prime)}_{\alpha \beta}$:
\begin{equation}
 \mathcal{P}^{(\prime)}_{\alpha \beta} = g_{\alpha \beta} - \frac{k_\alpha v^{(\prime)}_\beta+k_\beta v^{(\prime)}_\alpha}{v^{(\prime)} \cdot k}\, . \label{eq:kv-proj}
\end{equation}
Notice that $\breve{\varrho}^{(1)}$ again multiplies a soft factor, while the factors multiplied by $\breve{\varrho}^{(2,2')}$ are never singular as $k$ is light-like and $v^{(\prime)}$ time-like.

\subsection{Mesons}
\label{subsec:Mesons}
In this section, let us turn to study B-meson channels. 
We start from revisiting the $\bar{B}\to D^{(*)} l\bar{\nu} \gamma$ decay already analyzed in~\cite{Papucci:2021ztr}, in particular we propose a more convenient parameterization of the form factors, that makes much more straightforward the matching procedure onto \hhchpt.
Then, we focus on radiative semileptonic $D^{**}$ decays, where $D^{**}\in \qty{D_0^*, D_1^*, D_1, D_2^*}$ denotes the four lightest excited charmed mesons, above the HQS doublet $D^{1/2^-}\in\{D, D^*\}$. These four states are obtained combining the light quarks spins with a $l=1$ orbital angular momentum and they form two distinct HQS doublets
\begin{equation}
    D^{1/2^+}\in \qty{D_0^*, D_1^*}, \quad D^{3/2^+}\in\qty{D_1, D_2^*},
\end{equation}
where the spin in the exponent indicates the light quarks spin $s_q$ combined with the orbital angular momentum $l$. 
The analysis goes through along the same lines of the baryon case, the key difference being in the hadronic matrix elements, since in the meson case the colored light degrees of freedom carry half-integer angular momentum.

\subsubsection{$\Bar{B}\to D^{(*)} l \Bar{\nu}\gamma$ form factors}
\label{subsubsec: B to D(*) form factors}
The amplitudes for the $\bar{B}\to D^{*}$ transition are formally equal to the ones in \cref{eq:lepton matrix element,eq:light matrix element,eq:heavy matrix element} with the following substitutions:
\begin{equation}
    m_{\Lambda_b}\to m_{\Bar{B}}, \quad m_{\Lambda_c}\to m_{D^{(*)}},\quad F_\Gamma\to \Bar{F}_\Gamma, \quad M_\Gamma\to \Bar{M}_\Gamma, \quad G_{\Gamma, Q}\to \Bar{G}_{\Gamma,Q}.
\end{equation}
As already recalled above, in the HQS limit for the $\Bar{B}$ meson decay, the hadronic matrix element $\Bar{F}_\Gamma$ depend on a single Isgur-Wise function $\xi(w)$ ,
\begin{equation}
    \bar{F}_\Gamma=-\xi(w)\meHbHc{\Tr{\overline{\mathcal{H}}_{v'}^{(c)}\Gamma \mathcal{H}_v^{(b)}}}.
\end{equation}
At zero recoil $\xi(w)$ is normalized to unity.
As far as $\Bar{M}_\Gamma$, 
\begin{equation}
    \Bar{M}_{\Gamma}^{\mu}=\meHbHc{\Tr{\Bar{\mathcal{X}}^\mu\overline{\mathcal{H}}_{v'}^{(c)}\Gamma \mathcal{H}_v^{(b)}}},
\end{equation}
where\footnote{Although not immediately evident, the term proportional to $i\epsilon^{\mu\nu\rho\sigma}v_\nu v'_\rho k_\sigma\gamma^5$ can be reabsorbed in those already present in \cref{eq:X definition mesons} exploiting the properties of the super-fields summarized in \cref{app:hadronic super-fields}. Also the term proportional to $\epsilon^{\mu\nu\rho\sigma}v_\nu v'_\rho k_\sigma\slashed{k}\gamma^5$ is redundant, as it can be written in terms of other already present using the Chisholm identity.} 
\begin{align}
    \Bar{\mathcal{X}}^\mu&=\Bar{x}^{(1)}v^\mu+\Bar{x}^{(2)}v'^\mu+\Bar{x}^{(3)}\gamma^\mu+i\Bar{x}^{(4)}\sigma^{\mu\nu}k_\nu+ \Bar{x}^{(5)}v^\mu\slashed{k}+\Bar{x}^{(6)}v'^\mu\slashed{k},
\label{eq:X definition mesons}
\end{align}
$\sigma^{\mu\nu}=(i/2)[\gamma^\mu, \gamma^\nu]$ and $\epsilon^{0123}=1$. 
As above, we have dropped the terms proportional to $k^\mu$, as well as all the terms that do not transform as four-vectors under parity. The $\Bar{x}^{(j)}$ are scalar form factors depending on $w$, $v\cdot k$ and $v'\cdot k$. Their mass dimensions are $\qty[\Bar{x}^{(1-3)}]=-1$ and $\qty[\Bar{x}^{(4-6)}]=-2$. Compared to~\cite{Papucci:2021ztr}, we traded in the form factor basis the two operators $i\epsilon^{\mu\nu\rho\sigma}v_\nu v'_\rho k_\sigma\gamma^5$ and $(v+v')^\mu\slashed{k}$ for $v^\mu\slashed{k}$ and $v'^\mu\slashed{k}$. Although being less similar to the usual parameterization of non-radiative semileptonic decays, it is a convenient choice for two technical reasons: first, as we will discuss shortly, this parameterization holds also for the $D^{1/2^+}$ channel. Second, and perhaps more importantly, this choice of form factors (as it will be shown in \cref{subsec:sub-leading soft behavior}) renders the matching onto \hhchpt much more straightforward.
Finally, the hadronic matrix elements $\Bar{G}_\Gamma$ can be written as
\begin{equation}
    \bar{G}_{\Gamma,Q}=\meHbHc{\Tr{\bar{\mathcal{Y}}_Q\overline{\mathcal{H}}_{v'}^{(c)}\Gamma \mathcal{H}_v^{(b)}}},
\end{equation}
where
\begin{equation}
    \Bar{\mathcal{Y}}_Q=\Bar{y}^{(1)}_Q+\Bar{y}^{(2)}_Q\slashed{k}.
\end{equation}
The $\Bar{y}^{(j)}_Q$ are scalar form factors, whose mass dimensions are $\qty[\Bar{y}^{(1)}]=-1$ and $\qty[\Bar{y}^{(2)}]=-2$.

HQS constraints the number of new form factors, besides the Isgur-Wise function $\xi(w)$, down to ten: $\Bar{x}^{(1-6)}$ and $\Bar{y}^{(1-2)}_{b,c}$. Imposing the WI, that translates into the same additional constraints as in \cref{eq:Ward identity}, the computation gives\footnote{Here and again in the next subsection, in order to check explicitly, i.e. computing the trace for some specified $\Gamma$, one needs to use a convenient form of the Schouten's identity~\cite{Bondarev:1994ti} to reduce the expressions to the minimal number of independent terms involving the four dimensional Levi-Civita symbol, that is: $g_{\mu\nu} \epsilon_{\alpha\beta\lambda\rho}+g_{\mu\alpha} \epsilon_{\beta\lambda\rho\nu}+g_{\mu\beta} \epsilon_{\lambda\rho\nu\alpha}+g_{\mu\lambda} \epsilon_{\rho\nu\alpha\beta}+g_{\mu\rho} \epsilon_{\nu\alpha\beta\lambda}=0$.}
\begin{subequations}
    \begin{align}
        \bar{x}^{(2)}&=-\frac{v\cdot k}{v'\cdot k}\bar{x}^{(1)},\\
        \bar{x}^{(3)}&=-(v\cdot k) \bar{x}^{(5)}-(v'\cdot k) \bar{x}^{(6)},\\
        \bar{y}^{(1)}_b&=\frac{\xi}{v\cdot k},\\
        \bar{y}^{(1)}_c&=-\frac{\xi}{v'\cdot k},\\
        \bar{y}^{(2)}_{b,c}&=0.
    \end{align}
\end{subequations}
Thus, we remain with four new form factors other than the Isgur-Wise function $\xi$. The basis\footnote{The explicit relation between our basis of structure functions and the one presented in~\cite{Papucci:2021ztr} is: $\Bar{\varrho}^{(1)}=\zeta_1-(v\cdot k)(v'\cdot k)\zeta_3 $, $\Bar{\varrho}^{(3)}=\zeta_2-(w-1) \zeta_3$, $\Bar{\varrho}^{(2)}=\qty(v \cdot k)\qty(\zeta_3-\zeta_4)$, $\Bar{\varrho}^{(2')}=-\qty(v' \cdot k)\qty(\zeta_3 +\zeta_4)$, where $\zeta_i$ with $i=1,\dots,4$ are the form factors used there.} of such independent structure functions can be chosen to be
\begin{subequations}
    \begin{align}
        \Bar{\varrho}^{(1)}&\equiv (v\cdot k)\Bar{x}^{(1)},\\
        \Bar{\varrho}^{(2)}&\equiv -(v \cdot k)\Bar{x}^{(5)},\\
        \Bar{\varrho}^{(2')}&\equiv -(v' \cdot k)\Bar{x}^{(6)}, \\
        \Bar{\varrho}^{(3)}&\equiv \Bar{x}^{(4)}.
    \end{align}
\end{subequations}
The specific form of $\Bar{\varrho}^{(1)}$ anticipates the soft limit discussion of \cref{sec:soft limit}. $\Bar{\varrho}^{(1)}$ is dimensionless, while $\qty[\Bar{\varrho}^{(2,2')}] = -1$ and $\qty[\Bar{\varrho}^{(3)}] = -2$.  
With these choices and using \cref{eq:kv-proj}, the tensor $\bar{\mathcal{X}}^\mu$ becomes
    \begin{align}
        \bar{\mathcal{X}}^\mu=\bar{\varrho}^{(1)}\qty(\frac{v^\mu}{v\cdot k}-\frac{v'^\mu}{v'\cdot k})+\qty[\bar{\varrho}^{(2)}\mathcal{P}^\mu_{\alpha} +\bar{\varrho}^{(2')}\mathcal{P}^{\prime\mu}_\alpha] \gamma^\alpha +\bar{\varrho}^{(3)}i\sigma^{\mu\nu}k_\nu.
        \label{eq:XH}
    \end{align}

\subsubsection{$\Bar{B}\to D^{1/2^+} l \Bar{\nu}\gamma$ form factors}
The discussion for the $D^{1/2^+}$ doublet is very straightforward since this multiplet can be obtained by a parity transformation of the $D^{(*)}$ one. Therefore, the parameterization of the amplitude follows directly from the previous subsection. All the expressions match, upon performing the following substitutions
\begin{align}
    &m_{D^{(*)}}\to m_{D^{1/2^+}},\quad &&\Bar{F}_\Gamma \to \Tilde{F}_\Gamma, \quad &&&\Bar{M}_\Gamma \to \Tilde{M}_\Gamma, \quad &&&&\Bar{G}_{\Gamma, Q}\to \Tilde{G}_{\Gamma,Q},\nn\\* 
    &\xi(w)\to -2\tau_{1/2^+}(w),\quad &&\Bar{x}_Q^{(i)}\to\Tilde{x}_Q^{(i)},\quad &&&\Bar{y}_Q^{(i)}\to\Tilde{y}_Q^{(i)},\quad &&&&\Bar{\varrho}^{(i)}\to\Tilde{\varrho}^{(i)}.\quad\;
\end{align}
We have chosen the naming convention for the LO Isgur-Wise functions for the $D^{1/2^+}$ and $D^{3/2^+}$ multiplets of~\cite{Isgur:1990jf}, instead of e.g.~\cite{Leibovich:1997em,Leibovich:1997tu,Bernlochner:2017jxt}, as $\zeta(w)$ has already been used for the $\Lambda_c$ case.
The simplicity of the result stems from our choice of the form factor basis\footnote{Adopting the form factor basis of~\cite{Papucci:2021ztr} for this channel, the Dirac structures associated with $\Tilde{x}^{(5)}$ and $\Tilde{x}^{(6)}$ would be $i\epsilon^{\mu\nu\rho\sigma} v_\nu v'_\rho k_\sigma \gamma^5$ and $(v-v')^\mu\slashed{k}$ respectively, differing from the $D^{(*)}$ basis.}. Indeed, since parity is the only difference between $D^{(*)}$ and $D^{1/2^+}$ HQS multiplets, they only differ by a sign in the equation of motion and by the definition of the LO IW function, as detailed in \cref{app:hadronic super-fields}. Nevertheless, such a difference does not affect the chosen parameterization of the hadronic matrix elements.

\subsubsection{$\bar{B}\to D^{3/2^+} l \bar{\nu}\gamma$ form factors}
Finally, let us now consider the heaviest doublet $D^{3/2^+}$.
Again the amplitudes can be obtained from the ones in \cref{eq:lepton matrix element,eq:light matrix element,eq:heavy matrix element} with the following substitutions:
\begin{equation}
    m_{\Lambda_b}\to m_{\Bar{B}}, \quad m_{\Lambda_c}\to m_{D^{3/2^+}},\quad F_\Gamma\to \hat{F}_\Gamma, \quad M_\Gamma\to \hat{M}_\Gamma, \quad G_{\Gamma, Q}\to \hat{G}_{\Gamma,Q}.
\end{equation}
The matrix element $\hat{F}_\Gamma$ depends also in this case on a single Isgur-Wise function $\tau_{3/2}(w)$
\begin{equation}
    \hat{F}_\Gamma=\sqrt{3} \tau_{3/2^+}(w)\meHbFc{\Tr{v_\mu\overline{\mathcal{F}}_{v'}^{(c)\mu}\Gamma \mathcal{H}_v^{(b)}}},
\end{equation}
since $\gamma_\mu\overline{\mathcal{F}}_{v'}^\mu=0=v' \cdot\overline{\mathcal{F}}_{v'}$ and $\hat{F}_\Gamma$ cannot depend on $k_\mu$. As in the $\Lambda_{c1}^{(*)}$ case, the hadronic matrix element vanishes at zero recoil, and $\tau_{3/2}(1)$ does not have to be normalized to unity.
For $\hat{M}_\Gamma$ one has
\begin{equation}
    \hat{M}_{\Gamma}^{\mu}=\meHbFc{\Tr{\hat{\mathcal{X}}^\mu_{\,\sigma}\overline{\mathcal{F}}_{v'}^{(c)\sigma}\Gamma \mathcal{H}_v^{(b)}}},
\end{equation}
where\footnote{All the other terms, such as  $v_\sigma \epsilon^{\mu\rho\alpha\beta}v_\rho v'_\alpha k_\beta\gamma^5$ or $\epsilon^{\mu\nu\rho}_{\quad\;\,\sigma}v_\nu v'_\rho\gamma^5$, can be reabsorbed in those present in \cref{eq:Y tensor D(3/2)}.}
\begin{align}
    \hat{\mathcal{X}}^\mu_{\,\sigma}&=\hat{x}^{(1)}\delta^\mu_\sigma+\hat{x}^{(2)}v_\sigma v^\mu+\hat{x}^{(3)}v_\sigma v'^\mu+\hat{x}^{(4)}v_\sigma \gamma^\mu+\hat{x}^{(5)}k_\sigma v^\mu+\hat{x}^{(6)}k_\sigma v'^\mu+\hat{x}^{(7)}k_\sigma \gamma^\mu\nn\\
    &+\hat{x}^{(8)}\delta^\mu_\sigma\slashed{k}+\hat{x}^{(9)}v_\sigma v^\mu\slashed{k}+\hat{x}^{(10)}v_\sigma v'^\mu\slashed{k}+\hat{x}^{(11)}v_\sigma i\sigma^{\mu\nu}k_\nu \nn\\
    &+\hat{x}^{(12)}k_\sigma v^\mu\slashed{k}+\hat{x}^{(13)}k_\sigma v'^\mu\slashed{k}+\hat{x}^{(14)}k_\sigma i\sigma^{\mu\nu}k_\nu.
    \label{eq:Y tensor D(3/2)}
\end{align}
The $\hat{x}^{(j)}$ are scalar form factors and their mass dimensions are $\qty[\hat{x}^{(1-4)}]=-1$, $\qty[\hat{x}^{(5-11)}]=-2$ and $\qty[\hat{x}^{(12-14)}]=-3$.
Finally, the hadronic matrix elements $\hat{G}_\Gamma$ can be written as
\begin{equation}
    \hat{G}_{\Gamma,Q}=\meHbFc{\Tr{\hat{\mathcal{Y}}_{Q \mu}\overline{\mathcal{F}}_{v'}^{(c)\mu}\Gamma \mathcal{H}_v^{(b)}}},
\end{equation}
where
\begin{equation}
    \hat{\mathcal{Y}}_{Q \mu} =\hat{y}^{(1)}_Q v_\mu+\hat{y}^{(2)}_Q k_\mu+i \hat{y}^{(3)}_Q\epsilon_{\mu\nu\rho\sigma}v^\nu v'^\rho k^\sigma\gamma^5,
\end{equation}
since the terms proportional to $v'_\mu$ and $\gamma_\mu$ trivially vanish, the term $\sigma_{\mu\nu}k^\nu$ can be reabsorbed into the term proportional to $k_\mu$ and the term proportional to $v_\mu\slashed{k}$ can be written as a linear combination of the other three. The $\hat{y}^{(j)}_Q$ are scalar form factors depending on $w$, $v\cdot k$ and $v'\cdot k$, whose mass dimensions are $\qty[\hat{y}^{(1)}_Q]=-1$ and $\qty[\hat{y}^{(2-3)}_Q]=-2$.

Imposing the WI in \cref{eq:Ward identity}, we get
\begin{subequations}
    \begin{align}
        \hat{x}^{(1)}&=-(v\cdot k)\hat{x}^{(5)}-(v'\cdot k)\hat{x}^{(6)},\\
        \hat{x}^{(3)}&=-\frac{v\cdot k}{v'\cdot k}\hat{x}^{(2)},\\
        \hat{x}^{(4)}&=-(v\cdot k)\hat{x}^{(9)}-(v'\cdot k)\hat{x}^{(10)},\\
        \hat{x}^{(7)}&=-\hat{x}^{(8)}-(v\cdot k)\hat{x}^{(12)}-(v'\cdot k)\hat{x}^{(13)},\\
        \hat{y}^{(1)}_b&=-\sqrt{3}\frac{\tau_{3/2^+}}{v\cdot k},\\
        \hat{y}^{(1)}_c&=\sqrt{3}\frac{\tau_{3/2^+}}{v'\cdot k},\\
        \hat{y}^{(2)}_{b,c}&=\hat{y}^{(3)}_{b,c}=0,
    \end{align}
\end{subequations}
so that we are left with ten new independent structure functions.
The basis of such independent form factors can be chosen to be
\begin{subequations}
    \begin{align}
     \hat{\varrho}^{(1)}&\equiv (v\cdot k)\hat{x}^{(2)},\\
     \hat{\varrho}^{(2)}&\equiv -(v\cdot k)\hat{x}^{(9)},\\
     \hat{\varrho}^{(2')}&\equiv -(v'\cdot k)\hat{x}^{(10)},\\
     \hat{\varrho}^{(3)}&\equiv \hat{x}^{(11)},\\
     \hat{\varrho}^{(4)}&\equiv -(v\cdot k)\hat{x}^{(5)},\\
     \hat{\varrho}^{(4')}&\equiv -(v'\cdot k)\hat{x}^{(6)},\\
     \hat{\varrho}^{(5)}&\equiv \hat{x}^{(8)},\\
     \hat{\varrho}^{(6)}&\equiv -(v\cdot k)\hat{x}^{(12)},\\
     \hat{\varrho}^{(6')}&\equiv -(v'\cdot k)\hat{x}^{(13)},\\
     \hat{\varrho}^{(7)}&\equiv \hat{x}^{(14)}.
    \end{align}
\end{subequations}
Also in this case, the choice of $\hat{\varrho}^{(1)}$ is motivated by the soft limit, as it will be elucidated in the next section. The $\hat{\mathcal{X}}$ tensor become
\begin{align}\label{eq:XF}
    \hat{\mathcal{X}}^\mu_{\,\sigma}=&\hat{\varrho}^{(1)}\qty(\frac{v^\mu}{v\cdot k}-\frac{v'^\mu}{v'\cdot k})v_\sigma 
     +v_\sigma\qty[\hat{\varrho}^{(2)} \mathcal{P}^\mu_\alpha 
    +\hat{\varrho}^{(2')} \mathcal{P}^{\prime\mu}_\alpha] \gamma^\alpha + \hat{\varrho}^{(3)} \iu\sigma^{\mu\nu} k_\nu v_\sigma  \nn \\
    & +\qty[\hat{\varrho}^{(4)}\mathcal{P}^\mu_\sigma
    +\hat{\varrho}^{(4')}\mathcal{P}^{\prime \mu}_\sigma]+\hat{\varrho}^{(5)}\delta^\mu_{[\sigma} k_{\alpha]}   \gamma^\alpha+k_\sigma\qty[\hat{\varrho}^{(6)} \mathcal{P}^\mu_\alpha 
    +\hat{\varrho}^{(6')} \mathcal{P}^{\prime\mu}_\alpha] \gamma^\alpha +\hat{\varrho}^{(7)} \iu\sigma^{\mu\nu}k_\nu k_\sigma.
\end{align}
The mass dimension of the $\hat{\varrho}^{(i)}$ are: $\qty[\hat{\varrho}^{(1)}]=0$, $\qty[\hat{\varrho}^{(2,2',4,4')}]=-1$, $\qty[\hat{\varrho}^{(3,5,6,6')}]=-2$ and $\qty[\hat{\varrho}^{(7)}]=-3$.

\section{\hhchpt parameterization and soft limit}
\label{sec:soft limit}
For both baryons and mesons, further insight into the structure of the form factors can be achieved by studying the \emph{soft limit}, i.e. $k\ll \Lambda_{QCD}$. This limit is most easily accessed by matching HQET onto heavy hadron chiral perturbation theory (\hhchpt)~\cite{Wise:1992hn,Wise:1993wa,Burdman:1992gh,Cho:1992nt,Cho:1992cf,Cho:1992gg,Casalbuoni:1996pg}, describing the minimal coupling to the photon field~\cite{Cho:1992nt} and the higher order corrections due to the finite size of the heavy hadrons. 
\hhchpt keeps explicit the heavy quark and chiral symmetries, and it is organized in a derivative expansion, similarly to chiral perturbation theory, with a similar UV cutoff, $\Lambda_\chi \simeq 4\pi f_\pi$.

\subsection{Preliminaries} 
The leading order Lagrangian for the heavy hadron fields starts at $\mathcal{O}(p)$ and reads: 
\begin{align}
    \mathcal{L}^{(0)}_v& \supset \expval{\Hv(\iu v\cdot D)\Hvb}-\expval{\Kv(\iu v\cdot D)\Kvb}-g_{\mu\nu}\expval{\Fv^{\mu}(\iu v\cdot D)\Fvb^{\nu}}\nn\\
    &+\expval{\Tvb(\iu v\cdot \overleftarrow{D})\Tv}+g_{\mu\nu}\expval{\Rvb^\nu(i v\cdot \overleftarrow{D})\Rv^\mu}.
\end{align}
Here we have neglected the terms describing the interactions with the light mesons, not needed in the following. The angle brackets $\expval{\cdot}$ represent the sum over $SU(3)_V$ indices, and Dirac indices. Since we will not deal with QED transition amplitudes between different heavy multiplets, we have eliminated the large momentum component using the full multiplet mass. Alternatively, one could use a common field redefinition using only the lightest multiplets' masses ($\mathcal{H}$ and $\mathcal{T}$ for mesons and baryons respectively). In this case, there would be residual mass terms for the remaining multiplets.
All the heavy meson and baryon super-fields $\mathcal{H}$, $\mathcal{K}$, $\mathcal{F}$, $\mathcal{T}$, $\mathcal{R}$ transform as a $\bm{\bar{3}}$ of $SU(3)_V$. 
The covariant derivative in $\mathcal{L}^{(0)}_v$ acting on a $\bm{3}$ is~\cite{Cho:1992nt,Stewart:1998ke}:
\begin{equation}
D_\mu^{ab} =\delta^{ab}\partial_\mu-\iu \Vpi^{ab}_\mu+\iu e \qty(\mathcal{Q}^{ab} -e_Q \delta^{ab})A_\mu \nn
\end{equation}
with $\Vpi_\mu = \frac{\iu}{2}\qty(\xi^\dagger \mathcal{D}_\mu \xi + \xi \mathcal{D}_\mu \xi^\dagger)$, $\xi = \exp\qty(\iu \pi^a \lambda^a /f_\pi)$ and $\mathcal{D}_\mu \xi = \partial_\mu \xi + \iu e A_\mu \qty[\mathcal{Q}, \xi]$, with $\mathcal{Q}=diag(e_u, e_d,e_s)=diag(2,-1,-1)/3$. 

The well-known explicit expressions of the various super-fields are also reported in \cref{app:hadronic super-fields} for convenience.
Chiral symmetry breaking is introduced via the light quark mass operators $\Mpi_\pm = B_0 \qty( \xi^\dagger M_q \xi \pm \xi M_q \xi^\dagger)$, which are proportional to the light meson squared masses, $\mathcal{O}(p^2)$, and therefore not relevant for the rest of this paper. Further explicit breaking induced by the electromagnetic interactions is introduced via insertions of operators such as $ e \mathcal{Q}_\xi F_{\mu\nu}$ terms, with $F_{\mu\nu}$ the electromagnetic field strength and $\mathcal{Q}_\xi = \qty(\xi^\dagger \mathcal{Q} \xi + \xi \mathcal{Q}\xi^\dagger)/2$ with $\mathcal{Q}$ introduced above. All other dimension 2 operators would yield interactions with extra pions, not relevant here.

Since we will be focusing on the soft and next-to-soft regions, the $\mathcal{O}(p^2)$ Lagrangian describing higher-order interactions of the photons with heavy mesons will also be needed. It describes magnetic dipole transitions between heavy hadrons and reads\footnote{The term $\iu e F^{\mu\nu}\expval{\overline{\mathcal{F}}_\mu\mathcal{F}_\nu \mu_f}$ turns out to be proportional to the one present in \cref{eq:magnetic dipole contributions}.}~\cite{Cho:1992nt,Cho:1994vg}:
\begin{align}
    \mathcal{L}_{v}^{(1)}&\supset\frac{e}{2}F^{\mu\nu}\expval{\Hv \sigma_{\mu\nu} \mu_h \Hvb }
    +\frac{e}{2}F^{\mu\nu}\expval{\Kv \sigma_{\mu\nu}\mu_k \Kvb }\nn\\*
    &+\frac{e}{2}F^{\mu\nu}g_{\sigma\rho}\expval{ \Fv^{\sigma} \sigma_{\mu\nu} \mu_f \Fvb^{\rho}} 
    +\qty[\frac{\iu e}{2}F_{\mu\nu} \expval{ \Kv \gamma^\nu \mu_{fk}\Fvb^{\mu}}\plushc]\nn\\*
    &+\qty[\frac{e}{2}F^{\mu\nu}\expval{\Hv \sigma_{\mu\nu} \mu_{kh} \Kvb} \plushc ]
    +\qty[\frac{\iu e}{2}F_{\mu\nu}v^\nu \expval{\Hv \mu_{fh}\Fvb^\mu}\plushc] \nn\\*
    &+\frac{\iu e}{2}F_{\mu\nu}\expval{\Rvb^\mu  \mu_r\Rv^\nu}
    +\qty[\frac{\iu e}{2}F_{\mu\nu}v^\nu\expval{\Rvb^\mu  \mu_{rt}\Tv}\plushc],
    \label{eq:magnetic dipole contributions}
\end{align}
where $\mu_i= k_i\mathcal{Q}_\xi/\Lambda_\chi$, with $k_i~\mathcal{O}(1)$. There is no magnetic dipole operator for $\mathcal{T}$ as its light degrees of freedom are organized in a spinless configuration.

To describe the semileptonic decays, two copies of the previous Lagrangians, for the bottom and charm hadrons, with velocity labels $v$ and $v'$ respectively, are needed, together with the Lagrangian describing the weak current interactions:
\begin{align}
    \mathcal{L}_{W } &\supset  \frac{G_F V_{cb}}{\sqrt 2}  \sum_{\Gamma,\, \Gamma'} \alpha_{\Gamma \Gamma'} L_{\Gamma'} \circ  \mathcal{J}_\Gamma, \nn \\
    \mathcal{J}^{(0)}_\Gamma & = -\xi(w)\expval{\Hvpb  \Gamma \Hv} +2 \tau_{1/2^+}(w)\expval{ \Kvpb \Gamma \Hv} + \sqrt{3} \tau_{3/2^+} (w) v_{\mu}\expval{ \Fvpb^{\mu} \Gamma \Hv} \nn \\
    & ~ + \zeta(w)\expval{ \Tvpb \Gamma \Tv} + \sigma(w)v_{\mu}\expval{ \Rvpb^\mu \Gamma \Tv}, 
\end{align}

There are no $1/\Lambda_\chi$ corrections to the hadronic current involving photons at leading order in the heavy quark mass,  as we will show next in \cref{subsec:1overM}. This refines the results of~\cite{Papucci:2021ztr}: the operators in Eq.~39  in that paper only arise at $\mathcal{O}(1/m_Q)$.

\subsection{$1/m_Q$ corrections}
\label{subsec:1overM}
Violations of heavy quark symmetry, organized in a series of inverse powers of the heavy quark masses, are parameterized in \hhchpt by operators whose coefficients are suppressed by small parameters $\epsilon_{b,c} \sim \mathcal{O}(\Lambda/2 m_{b,c})$. Based on symmetry considerations, they can be divided into three different categories:
\begin{itemize}
\item Heavy quark flavor symmetry-violating, spin-symmetry preserving corrections: these contributions amount to shifting the coefficients of existing operators in the HQS symmetric \hhchpt Lagrangian by powers of $\epsilon_{b,c}$.
\item Operators constrained by reparameterization invariance (RPI)~\cite{Luke:1992cs}: in heavy particle effective theories, there is a residual symmetry stemming from the freedom of choice in splitting the heavy momenta in terms of a large component $m_Q v$ and a residual momentum $k \ll m_Q v$. The requirement that physics should not change under the transformation
\begin{equation}
k^\mu \rightarrow k^\mu + \varepsilon^\mu, \qquad v^\mu \rightarrow v^\mu - \frac{\varepsilon^\mu}{m_Q}
\end{equation}
links the coefficients of operators entering at different orders in the $1/m_Q$ expansion. As is well known~\cite{Neubert:1993iv,Heinonen:2012km}, the simplest way to identify such operators is to write the \hhchpt Lagrangian in terms of RPI-invariant super-fields and generalized velocities ($\mathcal{V}^\mu = v^\mu + i D^\mu/m_Q$) and then expand it in powers of the heavy mass. For convenience, we summarize the RPI-invariant super-fields in \cref{app:hadronic super-fields}. In practice, this procedure eliminates the possibility of having ``naked'' (i.e., not coming in an RPI-invariant combination) covariant derivatives acting directly on the heavy super-fields. Derivatives are still allowed on light degrees of freedom, which, for the purposes of this work it means we can use the photon field strength $F_{\mu\nu}$ together with the RPI-invariant building blocks. Since $F_{\mu\nu}$ is $\mathcal{O}(p^2)$, this statement automatically implies the absence of $\mathcal{O}(p)$ corrections to the hadronic current in the infinite mass limit.
\item Operators violating spin symmetry. These are genuine new operators whose coefficients is $1/m_{b,c}$ suppressed and are characterized by having a non-trivial Dirac structure connecting the heavy quark spin indices of the super-multiplets.
In particular, at $\mathcal{O}(1/m_{b,c})$, the new terms in the Lagrangian involving only the heavy fields, no pions and at most one photon read\footnote{Although in HQET the terms proportional to $\beta_2$ in Eq.~(14) of~\cite{Stewart:1998ke} are free parameters, they arise from the OPE between the $1/m$ Lagrangian with the light quark electromagnetic current. Investigating them in specific models, such as QCD sum rules \cite{Lu:2024tgy}, they tend to exhibit additional sources of suppression, e.g. $\lambda_{2,X}/\Lambda^2$. Such suppressions are also confirmed in the structure of the UV divergences for chiral loops~\cite{Stewart:1998ke} in \hhchpt for which such Lagrangian terms provide counterterms. Therefore, they may be considered together with $\mathcal{L}^{(2,1)}$.}
\begin{align}
\mathcal{L}^{(0,1)}_v   &\supset \frac{\lambda_{2,h}}{4 m_Q} \expval{\Hvb \sigma^{\mu\nu} \Hv \sigma_{\mu\nu}} + \frac{\lambda_{2,k}}{4 m_Q} \expval{\Kvb \sigma^{\mu\nu} \Kv \sigma_{\mu\nu}} + \frac{\lambda_{2,f}}{4 m_Q} g_{\rho\sigma}\expval{\Fvb^\rho \sigma^{\mu\nu} \Fv^\sigma \sigma_{\mu\nu}} \nn \\
& + \frac{\lambda_{2,kf}}{4 m_Q} \qty[\expval{\Kvb \sigma_{\mu\nu} \Fv^\mu \gamma^\nu} \plushc] + \frac{\lambda_{2,r}}{4 m_Q} \expval{\Rvb^\mu \sigma_{\mu\nu} \Rv^\nu}, \\
\mathcal{L}^{(1,1)}_v   &\supset-\frac{e e_Q}{4 m_Q}F^{\mu\nu}\expval{\Hvb \sigma_{\mu\nu} \Hv}
    -\frac{e e_Q}{4m_Q}F^{\mu\nu}\expval{\Kvb \sigma_{\mu\nu} \Kv } -\frac{e e_Q}{4m_Q}F^{\mu\nu}\expval{\overline{\mathcal{F}}_v^\sigma \sigma_{\mu\nu}\mathcal{F}_{v\sigma}}\nn\\
    &+\frac{e e_Q \tau_{1/2^+}(1)}{2m_Q} F^{\mu\nu}\qty[\expval{\overline{\mathcal{K}}_v  \sigma_{\mu\nu} \mathcal{H}_v} \plushc]+\frac{\iu e e_Q \sqrt{3} \tau_{3/2^+}(1)}{4m_Q}F_{\mu\nu}v^\nu \qty[\expval{\overline{\mathcal{F}}_v^\mu \mathcal{H}_v} \plushc] \nn \\
    &+\frac{\iu e e_Q \hat\xi''(1)}{4m_Q}F_{\mu\nu} \qty[\expval{ \overline{\mathcal{F}}_v^\mu \gamma^\nu \mathcal{K}_v} \plushc]
    -\frac{e e_Q }{4m_Q}F^{\mu\nu}\expval{\overline{\mathcal{T}}_v  \sigma_{\mu\nu}\mathcal{T}_v} \nn \\
    &-\frac{e e_Q }{4m_Q}F^{\mu\nu}\expval{\overline{\mathcal{R}}_v^\alpha  \sigma_{\mu\nu}\mathcal{R}_{v\alpha}} 
    +\frac{\iu e e_Q \sigma(1)}{4m_Q}F_{\mu\nu}v^\nu\qty[\expval{\overline{\mathcal{R}}_v^\mu  \mathcal{T}_v} \plushc],
    \label{eq:dipole contributions 1/M}
\end{align}
where we have used the notation $\mathcal{L}^{(a,b)}$ to indicate the Lagrangian at \mbox{N$^a$LO} in the \hhchpt derivative expansion and at $\mathcal{O}(1/m_Q^b)$ in the heavy quark mass expansion.
For the hadronic current, one has the standard $1/2m_{b,c}$ corrections matched onto \hhchpt. For example for the $\bar{B}\rightarrow D^{(*)}$ case one has~\cite{Boyd:1995pq,NPP}.
\begin{align}
    \mathcal{J}^{(0,1)}_\Gamma & \supset \rho_0(w)\qty(\varepsilon_b + \varepsilon_c)\expval{\Hvpb  \Gamma \Hv} + \rho_1(w)\qty(\varepsilon_b  \expval{\Hvpb   \Gamma \gamma^\mu \Hv \gamma_\mu} + \varepsilon_c \expval{\Hvpb  \gamma^\mu \Gamma \Hv \gamma_\mu}) \nn \\
    &\quad +\rho_2(w)\qty(\varepsilon_b  \expval{\Hvpb   \Gamma \slashed{v}' \Hv } + \varepsilon_c \expval{\Hvpb  \slashed{v} \Gamma \Hv \gamma_\mu}) \nn \\
    &\quad + \rho_3(w)\qty(\varepsilon_b  \expval{\Hvpb   \Gamma \Pi_+ \sigma^{\mu\nu} \Hv \sigma_{\mu\nu}} + \varepsilon_c \expval{\Hvpb  \sigma^{\mu\nu} \Pi'_+\Gamma \Hv \sigma_{\mu\nu}}) \nn \\
    & \quad+\iu\rho_4(w)\qty(\varepsilon_b  \expval{\Hvpb   \Gamma \Pi_+ \sigma^{\mu\nu} \Hv \gamma_{[\mu}v'_{\nu]}} - \varepsilon_c \expval{\Hvpb  \sigma^{\mu\nu} \Pi'_+\Gamma \Hv \gamma_{[\mu} v_{\nu]}})
\end{align}
where we have defined $\varepsilon_{b,c} = \Bar \Lambda / 2 m_{b,c}$ and the form factors $\rho_i(w)$ are linear combination of the usual sub-leading Isgur-Wise functions $\eta(w)$, $\chi_{1,2,3}(w)$.
\end{itemize}

\subsection{Leading soft behavior}
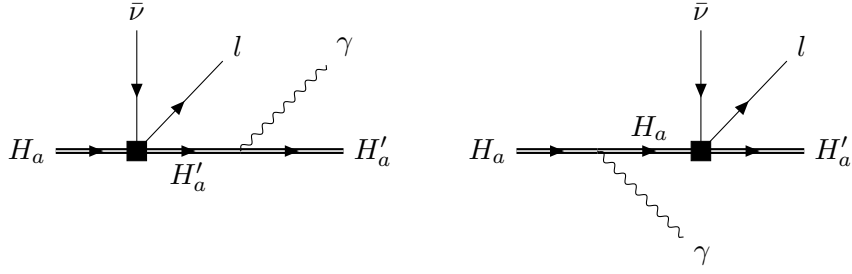
\begin{figure}       
    \centering
        \begin{tikzpicture}[baseline=(a.base)]
        \begin{feynman}[inline=(a.base)]
        \vertex (a) {\(H_a\)};
        \vertex [right= 1.45cm of a] (b);
        \vertex [right= 1.36cm of b] (i);
        \vertex [right = 1.36cm of i] (c) {\(H'_a\)};
        \vertex [above right= 1.6cm of b] (h) {\(l\)};
        \vertex [above = 1.6cm of b] (l) {\(\bar{\nu}\)};
        \vertex [above right= 1.6cm of i] (m) {\(\gamma\)};
        \diagram* {
        (a)  -- [double, thick, with arrow=0.5] (b) -- [double, thick, with arrow=0.5, edge label'=\(H'_a\)] (i) -- [double, thick, with arrow=0.5] (c);
        (b) -- [fermion] (h);
        (l) -- [fermion] (b);
        (i) -- [photon] (m);
        };
        \node at (b) [draw, fill];
        \end{feynman}
        \end{tikzpicture}
        \hspace{0.5cm}
        \begin{tikzpicture}[baseline=(a.base)]
        \begin{feynman}[inline=(a.base)]
        \vertex (a) {\(H_a\)};
        \vertex [right= 1.45cm of a] (i);
        \vertex [right= 1.36cm of b] (b);
        \vertex [right = 1.36cm of b] (c) {\(H'_a\)};
        \vertex [above right= 1.6cm of b] (h) {\(l\)};
        \vertex [above = 1.6cm of b] (l) {\(\bar{\nu}\)};
        \vertex [below right = 1.6cm of i] (m) {\(\gamma\)};
        \diagram* {
        (a)  -- [double, thick, with arrow=0.5] (i) -- [double, thick, with arrow=0.5, edge label=\(H_a\)] (b) -- [double, thick, with arrow=0.5] (c);
        (b) -- [fermion] (h);
        (l) -- [fermion] (b);
        (i) -- [photon] (m);
        };
        \node at (b) [draw, fill];
        \end{feynman}
        \end{tikzpicture}
        \caption{\hhchpt Feynman diagrams contributing to the leading soft behavior, i.e. $\mathcal{O}(k^{-1})$. The diagrams are the same either for the meson and for the baryon cases, so that here $H_a$ and $H'_a$ indicate two generic hadronic fields with $SU(3)_V$ index $a$. The black square indicates the weak current insertion.}
        \label{fig:leading order hhchpt diagrams}
\end{figure}

We are now in a position to study the infrared behavior of the amplitude $\mathcal{M}^\mu$ by expanding it in powers of the photon momentum $k$. The expansion starts at $k^{-1}$:
\begin{equation}
\mathcal{M}^\mu = \qty[\mathcal{M}^\mu]_{-1} + \qty[\mathcal{M}^\mu]_{0} + \dots
\end{equation}
where we have used the notation $\qty[O]_{i}$ to denote the $\mathcal{O}(k^i)$ term in the photon energy expansion of $O$.
As in~\cite{Papucci:2021ztr}, we will focus on the leading and next-to-leading soft terms, $\mathcal{O}(k^{-1})$ and $\mathcal{O}(k^0)$, the former being fixed by the QED soft photon theorem~\cite{Weinberg:1965nx}, while the latter providing the first corrections to the point-like limit of the heavy hadrons via non-vanishing contributions to the structure-dependent form factors. In principle one can continue this systematic expansion at any order in $k$, but the increasing number of unknown parameter, and/or functions will render this expansion impractical beyond the first few orders.

At leading order, there are two contributing diagrams, describing the photon emission from the hadronic line, as depicted in \cref{fig:leading order hhchpt diagrams}. The amplitude for a generic $B$-hadron radiative decay is
\begin{equation}
    \frac{\qty[\mathcal{M}^\mu]_{-1}}{\sqrt{m_{H_b} m_{H_c}}}=\frac{G_F e V_{cb}}{\sqrt{2}}\qty[\qty(\frac{(e_a-e_b)v^\mu}{v\cdot k}+\frac{(e_c-e_a)v'^\mu}{v' \cdot k})F_\Gamma]\bar{u}\Gamma' v+\frac{\qty[\mathcal{M}^\mu_{lepton}]_{-1}}{\sqrt{m_{H_b} m_{H_c}}},
\end{equation}
where $m_{H_b,H_c}$ denote the masses of the generic heavy hadrons and $\qty[\mathcal{M}^\mu_{lepton}]_{-1}$ is \cref{eq:lepton matrix element} expanded at $\mathcal{O}(k^{-1})$:
\begin{equation}
\frac{\qty[\mathcal{M}^\mu_{lepton}]_{-1}}{\sqrt{m_{H_b} m_{H_c}}} = e_l\frac{G_F e V_{cb}\alpha_{\Gamma \Gamma'}}{\sqrt{2}}\frac{F_\Gamma p_l^\mu}{2 p_l\cdot k}\bar{u} \gamma^\mu \Gamma' v.
\end{equation}
The strategy is now to match this expression onto the HQET matrix element expanded at the same order $k^{-1}$. This means we have to require
\begin{subequations}
    \begin{align}
        e_b v^\mu \qty[G_{\Gamma,b}]_{-1} &\stackrel{!}= -\frac{e_b v^\mu}{v\cdot k}F_\Gamma,\label{eq:first matching condition}\\
        e_c v'^\mu \qty[G_{\Gamma,c}]_{-1} &\stackrel{!}= \frac{e_c v'^\mu}{v'\cdot k}F_\Gamma,\label{eq:second matching condition}\\
        \qty[M_\Gamma^{\mu}]_{-1}&\stackrel{!}= e_a \qty(\frac{v^\mu}{v\cdot k}-\frac{v'^\mu}{v'\cdot k})F_\Gamma\label{eq:third matching condition}
    \end{align}
\end{subequations}
where the LHS is the HQET computation, while the RHS is the \hhchpt one.

\subsubsection{Baryons}
Let us start from considering the $\Lambda_b$ decays to $\Lambda_c$ and $\Lambda_{c1}^{(*)}$ baryons. The first two matching conditions \cref{eq:first matching condition,eq:second matching condition} are automatically satisfied. Third condition for the decay to the $\Lambda_c$ baryon one has
\begin{equation}\label{eq:rhomatchT0}
    \qty[\varrho^{(1)}]_0 = e_a \zeta(w) + \mathcal{O}\qty(\varepsilon_{b,c}).
\end{equation}
As expected, in soft limit, the form factors are controlled by the non-radiative Isgur-Wise function $\zeta(w)$.
Analogously, in the $\Lambda_{c1}^{(*)}$ channel, one has 
\begin{equation}\label{eq:rhomatchR0}
    \qty[\breve{\varrho}^{(1)}]_{0} = e_a \sigma(w) + \mathcal{O}\qty(\varepsilon_{b,c}), \quad \qty[\breve{\varrho}^{(2,2')}]_{-1} = 0.
\end{equation}
At this order $\breve{\varrho}^{(1)}$ is still fixed to be proportional the LO IW function $\sigma(w)$ while $\breve{\varrho}^{(2,2')}$ do not receive contributions at $\mathcal{O}(k^{-1})$. 
In both cases, the $1/m_Q$ contributions to $ \qty[\varrho^{(1)}]_{0}$, $ \qty[\breve{\varrho}^{(1)}]_{0}$ will depend on the sub-leading non-radiative IW functions. Although not proven here explicitly, it follows straightforwardly from the photon soft theorem that $\qty[\breve{\varrho}^{(2,2')}]_{-1}$ will not receive contributions at any order in the $1/m_Q$ expansion.

\subsubsection{Mesons}
Similarly to the baryon case, also for the $\Bar{B}$ decays to $D^{(*)}$ and $D^{**}$ mesons, the first two matching conditions, those containing the terms that describe the radiation of the photon from the heavy quarks, are automatically satisfied. As far as the third one, the result is analogous to the baryonic ones.
For the $D^{(*)}$, as already derived in~\cite{Papucci:2021ztr},
\begin{equation}\label{eq:rhomatchH0}
    \qty[\Bar{\varrho}^{(1)}]_{0}= -e_a \xi(w)+ \mathcal{O}\qty(\varepsilon_{b,c}), \quad \qty[\Bar{\varrho}^{(2,2')}]_{-1} = 0, \quad \qty[\Bar{\varrho}^{(3)}]_{-2} = 0.
\end{equation}
The $D^{1/2^+}$ case is identical, so that as expected
\begin{equation}\label{eq:rhomatchK0}
    \qty[\Tilde{\varrho}^{(1)}]_{0}= e_a 2 \tau_{1/2^+}(w)+ \mathcal{O}\qty(\varepsilon_{b,c}), \quad \qty[\Tilde{\varrho}^{(2,2')}]_{-1} = 0, \quad \qty[\Tilde{\varrho}^{(3)}]_{-2} = 0.
\end{equation}
Similarly, for the $D^{3/2^+}$ case, one has
\begin{align}\label{eq:rhomatchF0}
    \qty[\hat{\varrho}^{(1)}]_{0}&= e_a \sqrt{3} \tau_{3/2^+}(w)+ \mathcal{O}\qty(\varepsilon_{b,c}),  &\qty[\hat{\varrho}^{(2,2',4,4')}]_{-1} &= 0, \nn \\
     \qty[\hat{\varrho}^{(3,5,6,6')}]_{-2} &= 0,  &\qty[\hat{\varrho}^{(7)}]_{-3} &= 0.
\end{align}

\subsection{Sub-leading soft behavior}
\label{subsec:sub-leading soft behavior}
\begin{figure}
    \centering
    \begin{tikzpicture}[baseline=(a.base)]
        \begin{feynman}[inline=(a.base)]
        \vertex (a) {\(H_a\)};
        \vertex [right= 1.45cm of a] (b);
        \vertex [right= 1.36cm of b] (i);
        \vertex [right = 1.36cm of i] (c) {\(H'_a\)};
        \vertex [above right= 1.6cm of b] (h) {\(l\)};
        \vertex [above = 1.6cm of b] (l) {\(\bar{\nu}\)};
        \vertex [above right= 1.6cm of i] (m) {\(\gamma\)};
        \diagram* {
        (a)  -- [double, thick, with arrow=0.5] (b) -- [double, thick, with arrow=0.5, edge label'=\(H'^{(*)}_a\)] (i) -- [double, thick, with arrow=0.5] (c);
        (b) -- [fermion] (h);
        (l) -- [fermion] (b);
        (i) -- [photon] (m);
        };
        \node at (b) [draw, fill];
        \node at (i) [draw, fill, circle];
        \end{feynman}
        \end{tikzpicture}
        \hspace{0.5cm}
        \begin{tikzpicture}[baseline=(a.base)]
        \begin{feynman}[inline=(a.base)]
        \vertex (a) {\(H_a\)};
        \vertex [right= 1.45cm of a] (i);
        \vertex [right= 1.36cm of i] (b);
        \vertex [right = 1.36cm of b] (c) {\(H'_a\)};
        \vertex [above right= 1.6cm of b] (h) {\(l\)};
        \vertex [above = 1.6cm of b] (l) {\(\bar{\nu}\)};
        \vertex [below right = 1.6cm of i] (m) {\(\gamma\)};
        \diagram* {
        (a)  -- [double, thick, with arrow=0.5] (i) -- [double, thick, with arrow=0.5, edge label=\(H_a^{(*)}\)] (b) -- [double, thick, with arrow=0.5] (c);
        (b) -- [fermion] (h);
        (l) -- [fermion] (b);
        (i) -- [photon] (m);
        };
        \node at (b) [draw, fill];
        \node at (i) [draw, fill, circle];
        \end{feynman}
        \end{tikzpicture}
        \caption{\hhchpt Feynman diagrams of magnetic dipole operators contributing to the sub-leading soft behavior, i.e. $\mathcal{O}(k^0)$. Also in this case the diagrams for the meson and baryon case are the same, so that $H_a$ and $H'_a$ are two generic hadron fields with $SU(3)_V$ index $a$. The black square indicate the weak current insertion, while the black circle indicate the insertion of a dim-$5$ magnetic dipole operator.}
        \label{fig:higher order hhchpt diagrams}
\end{figure}
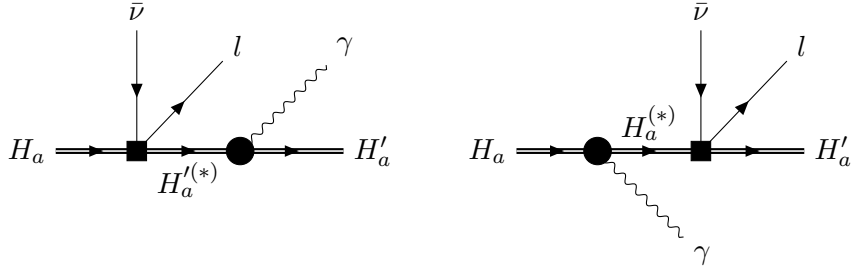

In order to study the sub-leading soft behavior, we have to match the HQET amplitude at $\mathcal{O}(k^0)$ to the \hhchpt one, this time including higher-order operators as in \cref{fig:higher order hhchpt diagrams}. By dimensional analysis, at sub-leading soft order, the photon emission will be controlled by \emph{magnetic dipole operators} in $\mathcal{L}_v^{(1)}$.

In the limit of small photon momentum, only dipole operators between states in the same multiplet will contribute. In fact, transitions between two distinct super-fields in a semileptonic decay diagram yield a propagator proportional to $1/\Delta M$, the mass splitting between multiplets, with $\Delta M/\Lambda_\chi$  generically not much smaller than unity (for the cases considered $\Delta M/\Lambda_\chi \sim 0.3 - 0.5$). Consequently, these contributions are formally suppressed at the order $\mathcal{O}\qty(1/\Lambda_\chi \Delta M)$ and will be considered higher order. There is, however, one exception where $\Delta M$ is subject to an accidental cancellation. Indeed, using the masses reported by the Particle Data Group (PDG)~\cite{pdg:pdg}, we find that $m_{D^{3/2^+}}-m_{D^{1/2^+}}\simeq 41$ MeV, $\Gamma_{D^{1/2^+}}/2\simeq 146$ MeV $\ll \Lambda_\chi$\footnote{Here, $m_{D^{3/2^+}}$, $m_{D^{1/2^+}}$, $\Gamma_{D^{1/2^+}}$ are the spin-averaged masses and widths of the corresponding HQ multiplets.}. Therefore, we will modify the naive power counting and consider the (complex) mass difference $\Delta M_{\mathcal{FK}}$ to be $\mathcal{O}(k)$, including $\mathcal{F} - \mathcal{K}$ dipole transitions and keeping the full dependence of the propagators in the corresponding diagrams.

Finally, we can write the \hhchpt amplitude for a generic B-hadron as
\begin{equation}
    \frac{\qty[\mathcal{M}^\mu]_{0}}{\sqrt{m_{H_b} m_{H_c}}}=\frac{G_F e V_{cb}}{\sqrt{2}} \Delta M_\Gamma^{\mu,(1)}\bar{u}\Gamma' v+\frac{\qty[\mathcal{M}^\mu_{lepton}]_0}{\sqrt{m_{H_b} m_{H_c}}},
\end{equation}
where $\qty[\mathcal{M}^\mu_{lepton}]_0$ is
\begin{equation}
\frac{\qty[\mathcal{M}^\mu_{lepton}]_0}{\sqrt{m_{H_b} m_{H_c}}} = \iu \, e_l\frac{e\, G_F V_{cb}\alpha_{\Gamma \Gamma'}}{\sqrt{2}}\frac{F_\Gamma }{2 p_l\cdot k}\bar{u} \sigma^{\mu\nu} k_\nu \Gamma' v.
\end{equation}
$\Delta M_{\Gamma}^{\mu,(1)}$ is the NLO \hhchpt hadronic amplitude, computed using \cref{eq:magnetic dipole contributions}. The matching conditions are straightforward, 
\begin{subequations}
    \begin{align}
        e_b v^\mu \qty[G_{\Gamma,b}^\nu]_0&\stackrel{!}=0,\label{eq:sub-soft matching 1}\\
        e_c v'^\mu \qty[G_{\Gamma,c}^\nu]_0&\stackrel{!}=0,\label{eq:sub-soft matching 2}\\
        \qty[M_\Gamma^{\mu}]_0&\stackrel{!}=\Delta M_\Gamma^{\mu,(1)},
    \end{align}
\end{subequations}
Given the fact that $G_{\Gamma, Q}$ is purely $\mathcal{O}(k^{-1})$, the conditions in \cref{eq:sub-soft matching 1,eq:sub-soft matching 2} are automatically satisfied.

\subsubsection{Baryons}
Since there is no magnetic dipole moment operator coupling two $\mathcal{T}$ fields in the HQS limit, we do not have any contribution at this order for the $\Lambda_b\to\Lambda_c l\bar{\nu}\gamma$ decay. Therefore, 
\begin{equation} \label{eq:rhomatchT1}
    \qty[\varrho^{(1)}]_1=\mathcal{O}\qty(\varepsilon_{b,c}k^1).
\end{equation}

As far as $\Lambda_b\to\Lambda_{c1}^{(*)} l\bar{\nu}\gamma$ we have the contribution from just one magnetic dipole operator, so that 
\begin{align}
    \Delta \breve{M}^{\mu,(1)}_\Gamma&=-\frac{i}{e}\frac{\sigma(w)\mathcal{O}_{r}^{\mu\alpha\sigma}(v')}{v'\cdot k}\Lambda'_{\sigma\rho}v^\rho \meTbRc{\overline{\mathcal{R}}^{(c)}_{v'\alpha}\Gamma\mathcal{T}^{(b)}_{v}}\nn \\
    &=\frac{\mu_r^a\sigma(w)}{2}\qty[- w \mathcal{P}^{\prime\mu\alpha}+ \frac{v\cdot k}{v'\cdot k} \mathcal{P}^{\mu\alpha}]\meTbRc{\overline{\mathcal{R}}^{(c)}_{v'\alpha}\Gamma\mathcal{T}^{(b)}_{v}},
\end{align}
where the projection tensor $\Lambda'_{\sigma\rho}=g_{\sigma\rho}-v'_\sigma v'_\rho$ comes from the $\mathcal{R}$ propagator numerator. The $\mathcal{O}_r$ tensor instead is defined in \cref{app:magnetic dipole moment operators}, where we collect all such tensors for the different operators in \cref{eq:magnetic dipole contributions}.
Matching this result to the $\breve{\varrho}^{(j)}$ form factors expansions, we get
\begin{subequations} \label{eq:rhomatchR1}
    \begin{align}
        \qty[\breve{\varrho}^{(1)}]_1&=\mathcal{O}\qty(\varepsilon_{b,c} \, k^1), \\
        \qty[\breve{\varrho}^{(2)}]_0&=\frac{\mu_r^a}{2}\frac{v \cdot k}{v' \cdot k}\sigma(w)+\mathcal{O}\qty(\varepsilon_{b,c} ),\\
        \qty[\breve{\varrho}^{(2')}]_0&=-\frac{\mu_r^a}{2} w\,\sigma(w)+\mathcal{O}\qty(\varepsilon_{b,c} ).
    \end{align}
\end{subequations}
As expected, being unconstrained in the exact soft limit, the magnetic dipole contributions to $\breve{\varrho}^{(2-3)}$ start at $\mathcal{O}(k^{0})$.

\subsubsection{Mesons}
Let us first consider the channel $\bar{B}\to D^{(*)}l\bar{\nu}\gamma$. In our form factor basis, the calculation provides
\begin{align}
    \Delta \bar{M}_\Gamma^{\mu,(1)}&=\frac{i}{e}\meHbHc{\Tr{\qty(\Pi_-'\frac{\mathcal{O}_{h}^\mu}{v'\cdot k}-\frac{\mathcal{O}_{h}^\mu}{v\cdot k}\Pi_-) \overline{\mathcal{H}}^{(c)}_{v'}\Gamma\mathcal{H}^{(b)}_{v}}}, 
\end{align}
where $\Pi_-,\Pi_-'=(1-\slashed{v})/2, (1-\slashed{v'})/2$. 
Let us denote the operator in parentheses as $\Bar{\mathbb{O}}^\mu$. Using the cyclicity of the trace and the equations of motion for the super-fields, we have
\begin{equation}
    \Bar{\mathbb{O}}^\mu=-\xi(w)\qty{\frac{\mu_h}{v'\cdot k}\qty[\iu \sigma^{\mu\nu} k_\nu+(v'\cdot k)\mathcal{P}^{\prime\mu}_\alpha \gamma^\alpha]-\frac{\mu_h}{v\cdot k}\qty[\iu \sigma^{\mu\nu} k_\nu-(v\cdot k)\mathcal{P}^{\mu}_\alpha \gamma^\alpha ]}.
\end{equation} 
With the latter equation, we can directly match the \hhchpt expression to the HQET parameterization, getting the contributions at $\mathcal{O}\qty(k^0/\Lambda_\chi)$ for the form factors defined in \cref{eq:X definition mesons}.
The matching provides
\begin{subequations}\label{eq:rhomatchH1}
    \begin{align}
        \qty[\Bar{\varrho}^{(1)}]_1&=\mathcal{O}(\varepsilon_{b,c} \, k^1), \\
        \qty[\Bar{\varrho}^{(2)}]_0&=-\mu_h^a\xi(w)+\mathcal{O}(\varepsilon_{b,c} ),\\
        \qty[\Bar{\varrho}^{(2')}]_0&=-\mu_h^a\xi(w)+\mathcal{O}(\varepsilon_{b,c} ), \\
       \qty[ \Bar{\varrho}^{(3)}]_{-1}&=\mu_h^a\qty(\frac{1}{v'\cdot k}-\frac{1}{v\cdot k})\xi(w)+\mathcal{O}(\varepsilon_{b,c} k^{-1}).
    \end{align}
\end{subequations}
We correct a factor 2 typo in~\cite{Papucci:2021ztr}: upon the change of basis provided in \cref{subsubsec: B to D(*) form factors}, our results coincide with the ones found there up to the transformation $\mu_h^a \to  2\mu_h^a$.

Turning to $\bar{B}\to D^{1/2^+}l\bar{\nu}\gamma$, we have
\begin{align}
    \Delta \Tilde{M}_\Gamma^{\mu}&=-\frac{i}{e}\meHbKc{\Tr{\Bigg(\Pi_+'\frac{\mathcal{O}_{k}^\mu}{v'\cdot k}+\frac{\mathcal{O}_{h}^\mu}{v\cdot k}\Pi_+\Bigg) \overline{\mathcal{K}}^{(c)}_{v'}\Gamma\mathcal{H}^{(b)}_{v}}}\nn\\
    &+\frac{i}{e}\meHbFc{\Tr{\Bigg(\Pi_-'\frac{\mathcal{O}^{\mu\nu}_{fk}}{v'\cdot k-\Delta M_{\mathcal{F}\mathcal{K}}} \Lambda'_{\nu\sigma}v^\sigma\Bigg)\overline{\mathcal{K}}^{(c)}_{v'}\Gamma\mathcal{H}^{(b)}_{v}}}.
\end{align}
The operator acting on the hadronic current can be simplified to
\begin{align}
    \Tilde{\mathbb{O}}^\mu&=2\tau_{1/2^+}(w)\qty{\frac{\mu^a_k}{v'\cdot k}\qty[-\iu\sigma^{\mu\nu}k_\nu-(v'\cdot k)\mathcal{P}^{\prime\mu}_\alpha \gamma^\alpha]+\frac{\mu^a_h}{v\cdot k}\qty[-i\sigma^{\mu\nu}k_\nu+(v\cdot k)\mathcal{P}^{\mu}_\alpha \gamma^\alpha]}\nn\\
    &-\frac{\sqrt{3}}{2}\tau_{3/2^+}(w)\frac{\mu_{fk}^a}{v'\cdot k -\Delta M_{\mathcal{F}\mathcal{K}}}\bigg\{w (v'\cdot k)\mathcal{P}^{\prime\mu}_\alpha \gamma^\alpha-(v\cdot k)\mathcal{P}^{\mu}_\alpha \gamma^\alpha\nn\\
    &+(v\cdot k)(v'\cdot k)\qty(\frac{v^\mu}{v\cdot k} - \frac{v'^\mu}{v'\cdot k})\bigg\}.
\end{align} 
Matching these results to the $\Tilde{\varrho}^{(j)}$ form factors, one gets 
\begin{subequations} \label{eq:rhomatchK1}
    \begin{align}
        \qty[\Tilde{\varrho}^{(1)}]_1&=-\frac{\sqrt{3}\mu_{fk}}{2}\frac{(v\cdot k)(v'\cdot k)}{v'\cdot k -\Delta M_{\mathcal{F}\mathcal{K}}}\tau_{3/2^+}(w)+\mathcal{O}(\varepsilon_{b,c} \, k^1), \\
        \qty[\Tilde{\varrho}^{(2)}]_0&=-\mu_h^a\tau_{1/2^+}(w)-\frac{\sqrt{3}\mu_{fk}}{2}\frac{ v\cdot k}{v'\cdot k -\Delta M_{\mathcal{F}\mathcal{K}}}\tau_{3/2^+}(w)+\mathcal{O}(\varepsilon_{b,c}),\\
        \qty[\Tilde{\varrho}^{(2')}]_0&=\mu_k^a\tau_{1/2^+}(w)+\frac{\sqrt{3} \mu_{fk}}{2}\frac{ w\,  v' \cdot k}{v'\cdot k -\Delta M_{\mathcal{F}\mathcal{K}}}\tau_{3/2^+}(w)+\mathcal{O}(\varepsilon_{b,c} ), \\
        \qty[\Tilde{\varrho}^{(3)}]_{-1}&=-\qty(\frac{\mu_h^a}{v\cdot k}+\frac{\mu_k^a}{v'\cdot k})\tau_{1/2^+}(w)+\mathcal{O}(\varepsilon_{b,c} \, k^{-1}).
    \end{align}
\end{subequations}
Finally, considering $\bar{B}\to D^{3/2^+}l\bar{\nu}\gamma$, we get
\begin{align}
    \Delta \hat{M}_\Gamma^{\mu}&=\frac{i}{e}\meHbFc{\Tr{\Bigg(\Pi_-'\frac{\mathcal{O}^{\mu\sigma\alpha}_{f}}{v'\cdot k}\Lambda'_{\alpha\rho}v^\rho-\frac{\mathcal{O}_{h}^{\mu}}{v\cdot k}\Pi_-v^\sigma\Bigg)\overline{\mathcal{F}}^{(c)}_{v'\sigma}\Gamma\mathcal{H}^{(b)}_{v}}}\nn\\
    &-\frac{i}{e}\meHbFc{\Tr{\Bigg(\Pi_+'\frac{\mathcal{O}^{\mu\sigma}_{fk}}{v'\cdot k+\Delta M_{\mathcal{F}\mathcal{K}}}\Bigg)\overline{\mathcal{F}}^{(c)}_{v'\sigma}\Gamma\mathcal{H}^{(b)}_{v}}}, 
\end{align}
where $\Delta M_{\mathcal{F}\mathcal{K}}\equiv m_{D^{3/2^+}}-m_{D^{1/2^+}}$.
The operator $ \hat{\mathbb{O}}^{\mu\sigma}$ can be simplified to
\begin{align}
    \hat{\mathbb{O}}^{\mu\sigma} =&-\sqrt{3}\tau_{3/2^+}(w)\qty{\frac{\mu^a_f}{v'\cdot k}\qty[\iu\sigma^{\mu\nu}k_\nu \slashed{k}+(v'\cdot k)\mathcal{P}^{\prime\mu}_\alpha \gamma^\alpha]-\frac{\mu^a_h}{v\cdot k}\qty[\iu\sigma^{\mu\nu}k_\nu-(v\cdot k)\mathcal{P}^{\mu}_\alpha \gamma^\alpha ]}v^\sigma\nn\\
    &+\tau_{1/2^+}(w)\frac{\mu^a_{fk}}{v'\cdot k +\Delta M_{\mathcal{F}\mathcal{K}}}\qty[(v'\cdot k)\mathcal{P}^{\prime\mu\sigma} + g^{\mu [\sigma} k^{\alpha]} \gamma_\alpha]
\end{align} 
Matching this expression onto the $\hat{\varrho}^{(j)}$ form factors, we get
\begin{subequations} \label{eq:rhomatchF1}
    \begin{align}
        \qty[\hat{\varrho}^{(1)}]_1&=\mathcal{O}(\varepsilon_{b,c}\, k^1), \\ 
        \qty[\hat{\varrho}^{(2)}]_0&=\mu_h^a\tau_{3/2^+}(w)+\mathcal{O}(\varepsilon_{b,c}),\\
        \qty[\hat{\varrho}^{(2')}]_0&=\mu_f^a\tau_{3/2^+}(w)+\mathcal{O}(\varepsilon_{b,c}),\\
        \qty[\hat{\varrho}^{(3)}]_{-1}&=\qty(\frac{\mu_f^a}{v'\cdot k}-\frac{\mu_h^a}{v\cdot k})\tau_{3/2^+}(w)+\mathcal{O}(\varepsilon_{b,c} k^{-1}),\\
        \qty[\hat{\varrho}^{(4)}]_0&=\mathcal{O}(\varepsilon_{b,c}),\\
        \qty[\hat{\varrho}^{(4')}]_0&=-\frac{\mu_{fk}}{2}\frac{v' \cdot k}{v'\cdot k+\Delta M_{\mathcal{F}\mathcal{K}}}\tau_{1/2^+}(w)+\mathcal{O}(\varepsilon_{b,c}),\\
        \qty[\hat{\varrho}^{(5)}]_{-1}&=\frac{1}{2}\frac{\mu_{fk}}{v'\cdot k+\Delta M_{\mathcal{F}\mathcal{K}}}\tau_{1/2^+}(w)+\mathcal{O}(\varepsilon_{b,c} k^{-1}),\\
        \qty[\hat{\varrho}^{(6,6')}]_{-1}&=\mathcal{O}(\varepsilon_{b,c}\,k^{-1}),\\
        \qty[\hat{\varrho}^{(7)}]_{-2}&=0.
    \end{align}
\end{subequations}
As expected $\hat{\varrho}^{(7)}$ does not receive $\mathcal{O}(k^{-2})$ contributions by dimensional analysis. More interestingly, also $\hat{\varrho}^{(6,6')}$ at $\mathcal{O}(k^{-1})$ and $\hat{\varrho}^{(4)}$ at $\mathcal{O}(k^0)$ do not receive corrections at NLO in \hhchpt. Nevertheless, they may generically receive contributions from $1/m_Q$ corrections\footnote{They may also receive contributions at higher order in $k$ from the dipole operators between different HQ multiplets that we disregarded.}.

\section{Discussion}
\label{sec:Summary}

In this paper we have continued the exploration of radiative semileptonic B-hadron decays initiated in~\cite{Papucci:2021ztr}. We have considered all the channels that could be, for different reasons, of phenomenological interest in the near future, both as direct experimental targets and as sources of systematic uncertainties to other measurements, such as $V_{cb}$ and lepton flavor universality (LFU) ratios $R(H_c)$. These include the processes $\bar{B}\to D^{**}l\nu \gamma$, $\Lambda_b \to \Lambda_{c}l\nu \gamma$ and $\Lambda_{c1}^{(*)} l\nu \gamma$, as well as $\bar{B}\to D^{(*)}l\nu \gamma$ already studied in~\cite{Papucci:2021ztr}.

Similar to the $\bar{B}\to D^{(*)} l\nu\gamma$ case, in all the decay modes studied, the stringent constraints dictated by HQS yield a $\mathcal{O}(10\times)$ reduction in the number of structure-dependent radiative form factors compared to the the general case where only Lorentz and the discrete symmetries of QCD are imposed. This is summarized in \cref{table:summary}, together with the definitions of the HQET SD functions.

Moving to study their infrared behavior by matching onto \hhchpt, we improve on~\cite{Papucci:2021ztr} by proving that reparameterization invariance fully constrains the sub-leading soft order behavior of the form factors to be proportional to the leading order non-radiative IW functions and a small set of couplings. This is a remarkable statement as no knowledge of new functions is needed, just of a few numbers controlling the magnetic dipole electromagnetic transitions among hadrons, shown in \cref{table:summary}. For the nine decay modes considered here, only five magnetic dipole coefficients suffice.
Out of these 5 numbers, one is well known, $\mu_h$, being fixed by the $D^*\to D\gamma$ ($B^*\to B\gamma$) partial branching fraction~\cite{pdg:pdg,Stewart:1998ke}. Another one, $\mu_k$ is determined, albeit with sizable uncertainties, by the measurement of the branching ratio of $D_{s1} \to D_s \gamma$~\cite{pdg:pdg}. There are also upper bounds on the branching ratios of $D_{s0}^*\to D_s^*\gamma$, $D_{s2}^* \to D_s\gamma$ and $\Lambda_c^* \to \Lambda_c \gamma$ which constrain the maximum size of $\mu_{f}$ and $\mu_r$. Finally, $\mu_{fk}$ is currently unconstrained. Nevertheless, further information may be gained with dedicated measurements of electromagnetic decay modes $H_c' \to H_c \gamma$, with $H_c^{(\prime)}$ the modes of interest here, at Belle II and BES III.

This concludes the control of the photon spectrum for the real emission in semileptonic heavy to heavy decays up to corrections of $\mathcal{O}(E_\gamma/\Lambda_\chi)$ relative to the soft photon approximation. For a photon of $E_\gamma\sim 100\,{\rm MeV}$, above threshold for Belle II but potentially challenging for LHCb, corrections of $\mathcal{O}(15\%)$ are expected to the photon emission rate from hadronic legs, and additional photon emissions out of neutral hadrons, as they do not radiate in the soft photon approximation. For completeness, we mention that all other relevant QED effects have been already considered in the literature. Notably, Sudakov-type logarithms are important and they are already resummed in the eikonal approximation in state of the art Monte Carlo simulations \cite{ Barberio:1990ms, Barberio:1993qi, Sch_nherr_2008, Skands_2020}. Coulomb enhancements \cite{deBoer:2018ipi} and virtual corrections \cite{Becirevic:2009fy,Tostado_2016} are also discussed in the literature. The former still need to be incorporated into event generators, while the latter are generally sub-leading.

It is interesting to discuss the next corrections to the structure dependent effects we calculated to asses their robustness. The largest of these are the  $\mathcal{O}(\Lambda/2m_c)$ corrections. 
However, as we have explained in~\cref{subsec:1overM}, in the soft and sub-leading soft regions these do not introduce any more parameters and are fully calculable in terms of the parameters discussed above and non-radiative sub-leading (in $1/2m_{b,c}$) Isgur-Wise functions. Even the spin-symmetry conserving flavor-symmetry violating non-RPI corrections that shifts the \hhchpt couplings by $\mathcal{O}(\epsilon_{b,c})$ amounts, can either be reabsorbed in a redefinition of the LO couplings, or, in the case of the $\mathcal{H}_{b,c}$ multiplets are fully determined by the knowledge of both $B^* \to B\gamma$ and $D^*\to D\gamma$ partial width. The only source of uncertainties is the fact that certain sub-leading IW functions, which can be reabsorbed into the LO ones in the non-radiative case~\cite{Leibovich:1997em,Leibovich:1997tu}, may become physical in experimentally accessible decay modes once the photon emission is included. We will leave this question to future study.

\begin{table}[t]
\begin{adjustbox}{center}
\begin{tabular}{||c||x{1.8cm}||x{2.9cm}|c||x{2.5cm}|x{2.5cm}||}
\hline
Mode&\# of SD functions&HQET SD FF & $\mathcal{X}^\mu_{(\sigma)}$&\hhchpt $\mathcal{O}(k^0)$ parameters&Matching HQET/\hhchpt\\
\hline
$\Lambda_b \to \Lambda_c$ & $\mathcal{O}(10)$ & $ \rho^{(1)}$ & \eqref{eq:XT} & none & \eqref{eq:rhomatchT0},\eqref{eq:rhomatchT1} \\
$\Lambda_b \to \Lambda_{c1}^{(*)}$ & $\mathcal{O}(50)$ & $\breve \rho^{(1,2^{(\prime)})}$ & \eqref{eq:XR} & $\mu_{r}^a$ & \eqref{eq:rhomatchR0},\eqref{eq:rhomatchR1} \\
$\bar{B} \to D^{(*)}$ & 8+24 & $\Bar \rho^{(1,2^{(\prime)},3)}$ & \eqref{eq:XH} & $\mu_h^a$ & \eqref{eq:rhomatchH0},\eqref{eq:rhomatchH1} \\
$\bar{B} \to \qty(D_0^*, D_1^*)$ & 8+24 & $\tilde \rho^{(1,2^{(\prime)},3)}$ & \eqref{eq:XH} & $\mu_{fk}^a$, $\mu_h^a$, $\mu_{k}^a$ & \eqref{eq:rhomatchK0},\eqref{eq:rhomatchK0} \\
$\bar{B} \to \qty(D_1, D_2^*)$ & 24+$\mathcal{O}(50)$ & $\hat \rho^{(1,2^{(\prime)},3,4^{(\prime)},5,6^{(\prime)},7)}$ & \eqref{eq:XF} & $\mu_{fk}^a$, $\mu_h^a$, $\mu_{f}^a$ & \eqref{eq:rhomatchF0},\eqref{eq:rhomatchF1} \\
\hline
\end{tabular}
\end{adjustbox}
\caption{Summary table of the results of this work. For each decay mode, the list structure-dependent Isgur-Wise form factors in the HQS limit is given, together with the reference to their definitions. As a comparison, the number of general structure-dependent functions is also provided (or estimated). The last two columns list the \hhchpt magnetic dipole parameters controlling the next-to-soft limit of the SD functions and the corresponding HQET -- \hhchpt matching relations. The number of SD functions in the general case has only been estimated for $\Lambda_b\to \Lambda_c, \Lambda_{c1}^{(*)}$ and $\bar{B}\to D^{3/2^+}$, as the calculation would not be particularly illuminating.
}
\label{table:summary}
\end{table}

The next corrections are going to be suppressed by $\mathcal{O}(1/\Lambda_\chi^2)$ and $\mathcal{O}(1/\Lambda_\chi \Delta M)$. At this order they will introduce truly new structure dependent form factors, in the form of new IW functions multiplying contact operators (at $\mathcal{O}(1/\Lambda_\chi^2)$), or LO non-radiative IW functions of semileptonic decay modes not accessible experimentally (at $\mathcal{O}(1/\Lambda_\chi \Delta M)$ via diagrams similar to those of \cref{fig:higher order hhchpt diagrams} but with propagators introducing an extra $\Delta M$ suppression). They are listed in \cref{app:additional Isgur-Wise functions}. For a $\sim 100$~MeV photon, the latter contributions are expected to introduce corrections of $\mathcal{O}(E_\gamma/\Delta M)\sim 30\%$ to the sub-leading soft corrections computed in this paper, and determine the level uncertainties in the results presented here.

We report the full matrix elements for the SM currents in \cref{app:ME}. In order to render the results we derived here usable by the experimental collaborations, they should be included in event generators, as they are not captured by current simulation packages. This can be achieved with event reweighing tools such as \texttt{HAMMER}~\cite{Bernlochner:2020tfi}. We defer the \texttt{HAMMER} implementation of our expressions, in the form of helicity amplitudes, including also the calculable $1/m_c$ corrections, together with a study of their phenomenological implications to an upcoming publication~\cite{CP}.  
 
\acknowledgments
We thank Mark Wise for useful conversations. We used \texttt{FeynCalc}~\cite{Shtabovenko:2020gxv, Mertig:1990an} to check most of our analytical results and \texttt{TikZ-Feynman}~\cite{Ellis:2016jkw} to draw all the Feynman diagrams in this paper. FC and MP are supported by the U.S. Department of Energy, Office of Science, Office of High Energy Physics under Award Number DE-SC0011632, and by the Walter Burke Institute for Theoretical Physics. MP would like to thank the Aspen Center for Physics, which is supported by National Science Foundation grant PHY-2210452, where part of this work was performed.

\appendix
\section{Hadronic super-fields}
\label{app:hadronic super-fields}
For convenience we summarize the super-field expressions for the HQS multiplets used in this paper~\cite{Falk:1991nq}. 

Baryons with light quark spin $s_q=0$ and vanishing orbital angular momentum $l=0$ (e.g. $\Lambda_c$ and $\Lambda_b$ baryons) are described by 
\begin{subequations}
    \begin{align}
        \mathcal{T}_v=\frac{1+\slashed{v}}{2}T_v,\\
        \overline{\mathcal{T}}_v=\overline{T}_v\frac{1+\slashed{v}}{2},
    \end{align}
\end{subequations}
where $T_v$ is a Dirac spinor.
Baryon doublets with $s_q=0$ and $l=1$ (e.g. $\Lambda_{c1}^{(*)}$ baryons) are described by 
\begin{subequations}
    \begin{align}
        \mathcal{R}_v^\mu&=\sqrt{\frac{1}{3}}(\gamma^\mu +v^\mu)\gamma^5\frac{1+\slashed{v}}{2}R_v+\frac{1+\slashed{v}}{2}R^{*\mu}_v,\\
         \overline{\mathcal{R}}_v^\mu&=-\sqrt{\frac{1}{3}} \overline{R}_v \frac{1+\slashed{v}}{2}\gamma^5(\gamma^\mu +v^\mu)+\overline{R}^{*\mu}_v\frac{1+\slashed{v}}{2},
    \end{align}
\end{subequations}
where $R_v$ is a Dirac spinor, while $R_v^{*\mu}$ is a Rarita-Schwinger field satisfying the constraints $\partial_\mu R_v^{*\mu}= \gamma_\mu R_v^{*\mu}=0$.

Meson doublets with light quarks spin $s_q=1/2$ and vanishing orbital angular momentum $l=0$ (e.g. $\Bar{B}^{(*)}$ and $D^{(*)}$ mesons) are described by:
\begin{subequations}
    \begin{align}
        \mathcal{H}_v&=\frac{1+\slashed{v}}{2}\qty[-B_v \gamma^5+B^*_v \slashed{\epsilon}],\\
        \overline{\mathcal{H}}_v&=\qty[B_v^\dagger \gamma^5 +B^{* \dagger}_v \slashed{\epsilon^*}]\frac{1+\slashed{v}}{2},
    \end{align}
\end{subequations}
where $B_v$ and $B_v^*$ are the pseudo-scalar and vector wave-functions respectively.
$\mathcal{H}_v$ satisfies $\slashed{v}\mathcal{H}_v=\mathcal{H}_v$ and $\mathcal{H}_v\slashed{v}=-\mathcal{H}_v$.
Meson doublets with $s_q=1/2$ and $l=1$ (e.g. $D^{1/2^+}$ mesons) are described by:
\begin{subequations}
    \begin{align}
        \mathcal{K}_v&=\frac{1+\slashed{v}}{2}\qty[P_v+V_v \gamma^5 \slashed{\epsilon}],\\
        \overline{\mathcal{K}}_v&=\qty[P_v^\dagger-V^\dagger_v \slashed{\epsilon^*}\gamma^5 ]\frac{1+\slashed{v}}{2},
    \end{align}
\end{subequations}
where $P_v$ and $V_v^*$ are the scalar and pseudo-vector wave-functions respectively.
$\mathcal{K}_v$ satisfies $\slashed{v}\mathcal{K}_v=\mathcal{K}_v$ and $\mathcal{K}_v\slashed{v}=\mathcal{K}_v$.
Meson doublets with $s_q=3/2$ and $l=1$ (e.g. $D^{3/2^+}$ mesons) are described by:
\begin{subequations}
    \begin{align}
        \mathcal{F}_v^\mu &=\frac{1+\slashed{v}}{2}\qty{T_v\epsilon^{\mu\nu}\gamma_\nu-V_v\sqrt{\frac{3}{2}}\gamma^5 \qty[\epsilon^\mu -\frac{1}{3}\slashed{\epsilon}(\gamma^\mu-v^\mu)]},\\
        \overline{\mathcal{F}}_v^\mu &=\qty{T_v^\dagger\epsilon^{*\mu\nu}\gamma_\nu+V_v^\dagger\sqrt{\frac{3}{2}} \qty[\epsilon^{*\mu} -\frac{1}{3}(\gamma^\mu-v^\mu)\slashed{\epsilon^*}]\gamma^5}\frac{1+\slashed{v}}{2},
    \end{align}
\end{subequations}
where $\epsilon^\mu$ and $\epsilon^{\mu\nu}$ are the pseudo-vector and tensor polarization vectors respectively. $\epsilon^{\mu\nu}$ is symmetric and traceless.
$\mathcal{F}_v^\mu$ satisfies $\slashed{v}\mathcal{F}^\mu_v=\mathcal{F}^\mu_v$, $\mathcal{F}^\mu_v\slashed{v}=-\mathcal{F}^\mu_v$, $\mathcal{F}^\mu_v\gamma_\mu=0$ and $v\cdot\mathcal{F}_v=0$. 

Finally, we also report their RPI-invariant combinations at $\mathcal{O}(1/M)$, both for the baryons
\begin{subequations}
    \begin{align}
        \mathcal{T}&\to \mathcal{T}+\frac{i}{2 m_\mathcal{T}}\slashed{D}\mathcal{T},\\
        \mathcal{R}^\mu&\to \mathcal{R}^\mu+\frac{i}{2 m_\mathcal{R}}\slashed{D} \mathcal{R}^\mu-v^\mu \frac{i D\cdot \mathcal{R}}{m_\mathcal{R}},\\
    \end{align}
\end{subequations}
and for the mesons
\begin{subequations}
    \begin{align}
        \mathcal{H}&\to \mathcal{H}+\frac{i}{2 m_\mathcal{H}}[\slashed{D}, \mathcal{H}],\label{eq:H RPI invariant}\\
        \mathcal{K}&\to \mathcal{K}+\frac{i}{2 m_\mathcal{K}}\{\slashed{D}, \mathcal{K}\},\\
        \mathcal{F}^\mu&\to \mathcal{F}^\mu+\frac{i}{2 m_\mathcal{F}}[\slashed{D}, \mathcal{F}^\mu]-v^\mu\frac{i D\cdot \mathcal{F}}{m_\mathcal{F}}.
    \end{align}
\end{subequations}

\section{Hadronic matrix elements}
\label{app:ME}
We report the explicit expressions of the SM hadronic matrix elements that were not detailed in the main body. We strip the photon polarization, with Lorentz index $\mu$, and write the spin-1 and 2 hadron polarization vectors and tensors, $\epsilon^\alpha$, $\epsilon^{\alpha\beta}$, without any additional subscripts. We report these results in the HQET parameterization, in the HQS limit after imposing the WI. We will use the definition \cref{eq:kv-proj} of the projectors $\mathcal{P}^{(')}_{\alpha\beta}$ and the short-hand
\begin{equation}
\mathcal{S}^\mu = \qty(\frac{v^\mu}{v \cdot k} - \frac{v^{\prime \mu}}{v' \cdot k})
\end{equation}
for the soft factor. 
Finally, we adopt the following convention for the Levi-Civita tensor: $Tr[\gamma^\mu \gamma^\nu \gamma^\rho\gamma^\sigma\gamma^5]=4i\epsilon^{\mu\nu\rho\sigma}$.

\subsection{$\Bar{B}\to D l\Bar{\nu}\gamma$}
\begin{subequations}
    \begin{align}
    \Bar{F}_V^\nu &=  \xi (v+v')^\nu,\\ 
    i\Bar{F}_A^\nu &= 0.
\end{align}
\end{subequations}

\begin{subequations}
    \begin{align}
    \Bar{M}^{\mu\nu}_V&=-\Bar{\varrho}^{(1)} \mathcal{S}^\mu\,(v+v')^\nu+\Bar{\varrho}^{(3)}\qty[(v'\cdot k)\mathcal{P}^{\prime\mu\nu} -(v\cdot k)\mathcal{P}^{\mu\nu}]\nn\\
    &-\Bar{\varrho}^{(2)} [(w-1)\mathcal{P}^{\mu\nu}-v'^\alpha \mathcal{P}^{\mu}_{\alpha} v^\nu]
    -\Bar{\varrho}^{(2')}\qty[(w-1)\mathcal{P}^{\prime\mu\nu}-v^\alpha \mathcal{P}^{\prime\mu}_\alpha v'^\nu], \\     
    -\iu \Bar{M}^{\mu\nu}_A&=\Bar{\varrho}^{(3)} \epsilon^{\mu\nu\rho\sigma}k_\rho(v+v')_\sigma- \epsilon^{\alpha\nu\sigma\rho} v_\sigma v'_\rho \qty[\Bar{\varrho}^{(2)}\mathcal{P}^\mu_\alpha + \Bar{\varrho}^{(2')} \mathcal{P}^{\prime\mu}_\alpha].
\end{align}
\end{subequations}

\begin{subequations}
    \begin{align}
    &\Bar{G}^\nu_{V,c}=\frac{\xi}{v'\cdot k}(v+v')^\nu, && \Bar{G}^\nu_{V,b}=-\frac{\xi}{v\cdot k}(v+v')^\nu,\\ 
    &\Bar{G}^\nu_{A,c}=0,&& \Bar{G}^\nu_{A,b}=0.
\end{align}
\end{subequations}
\newline

\subsection{$\Bar{B}\to D^* l\Bar{\nu}\gamma$}
\begin{align}
    \Bar{F}_V^\nu &= -\iu\xi \epsilon^{\nu\alpha\rho\sigma}v_\alpha v'_\rho \epsilon^*_\sigma,\\ 
    \Bar{F}_A^\nu &= \xi [(w+1)\epsilon^{*\nu}-(v\cdot\epsilon^*)v'^\nu].
\end{align}

\begin{subequations}
    \begin{align}
    -\iu\Bar{M}^{\mu\nu}_V&=\Bar{\varrho}^{(1)}\mathcal{S}^\mu\epsilon^{\nu\alpha\rho\sigma}v_\alpha v'_\rho \epsilon^*_\sigma + \epsilon^{\alpha\nu\rho\sigma}\qty(v'-v)_\rho\epsilon^*_\sigma \qty[\Bar{\varrho}^{(2)}\mathcal{P}^\mu_\alpha +\Bar{\varrho}^{(2')} \mathcal{P}^{\prime\mu}_\alpha]\nn \\
    &\Bar{\varrho}^{(3)}\big[\epsilon^{*\mu}\epsilon^{\nu\alpha\rho\sigma}v_\alpha v'_\rho k_\sigma+ \epsilon^{\mu\alpha\rho\sigma} k_\rho \epsilon^*_\sigma(v'^\nu v_\alpha+v^\nu v'_\alpha) -\epsilon^{\mu\nu\rho\sigma}[(w-1)k_\rho \epsilon^*_\sigma-(k\cdot \epsilon^*)v_\rho v'_\sigma]\big]\nn\\
    \Bar{M}^{\mu\nu}_A&=-\Bar{\varrho}^{(1)}\mathcal{S}^\mu\qty[(w+1)\epsilon^{*\nu}-(v\cdot \epsilon^*)v'^\nu]\nn\\
    &+\Bar{\varrho}^{(3)} \big\{(w+1)\epsilon^*_\alpha g^{\mu[\nu} k^{\alpha]} +v^{[\mu}v^{\prime\nu]} \qty(k \cdot \epsilon^*) - \qty( v' \cdot k)\qty(v \cdot \epsilon^*) g^{\mu\nu}\nn\\
    &\qquad \quad +\epsilon^{*\alpha}[(v'\cdot k)\mathcal{P}^{\prime\mu}_\alpha v^\nu+(v\cdot k)\mathcal{P}^\mu_\alpha v'^\nu ]+\epsilon^{*\alpha}[(v'\cdot k)\mathcal{P}^{\prime\nu}_\alpha v^\mu-(v\cdot k)\mathcal{P}^\nu_\alpha v'^\mu]\big\}\nn\\
    &-\Bar{\varrho}^{(2)}\qty{(v\cdot\epsilon^*)\mathcal{P}^{\mu\nu}-\mathcal{P}^\mu_\alpha\epsilon^{*\alpha}(v-v')^\nu-\mathcal{P}^\mu_\alpha(v+v')^\alpha \epsilon^{*\nu}}\nn\\
    &-\Bar{\varrho}^{(2')}\qty{(v\cdot\epsilon^*)\mathcal{P}^{\prime\mu\nu}-\mathcal{P}^{\prime\mu}_\alpha\epsilon^{*\alpha}(v-v')^\nu-\mathcal{P}^{\prime\mu}_\alpha(v+v')^\alpha \epsilon^{*\nu}}.
\end{align}
\end{subequations}

\begin{subequations}
    \begin{align}
    &\Bar{G}^\nu_{V,c}=-\frac{\xi}{v'\cdot k}\iu\epsilon^{\nu\alpha\rho\sigma}v_\alpha v'_\rho \epsilon^*_\sigma, &&\Bar{G}^\nu_{V,b}=\frac{\xi}{v\cdot k}\iu\epsilon^{\nu\alpha\rho\sigma}v_\alpha v'_\rho \epsilon^*_\sigma, \\
    &\Bar{G}^\nu_{A,c}=\frac{\xi}{v'\cdot k}[(w+1)\epsilon^{*\nu}-(v\cdot\epsilon^*)v'^\nu], &&\Bar{G}^\nu_{A,b}=-\frac{\xi}{v\cdot k}[(w+1)\epsilon^{*\nu}-(v\cdot\epsilon^*)v'^\nu].
\end{align}
\end{subequations}
\newline

\subsection{$\Bar{B}\to D_0^* l\Bar{\nu}\gamma$}
\begin{subequations}
    \begin{align}
    \Tilde{F}_V^\nu &= 0,\\ 
    \Tilde{F}_A^\nu &= 2 \tau_{1/2} (v-v')^\nu.
\end{align}
\end{subequations}

\begin{subequations}
    \begin{align}
    -\iu\Tilde{M}^{\mu\nu}_V&=\epsilon^{\nu\alpha\rho\sigma}v_\rho v'_\sigma \qty[\Tilde{\varrho}^{(2)} \mathcal{P}^\mu_\alpha + \Tilde{\varrho}^{(2')} \mathcal{P}^{\prime\mu}_\alpha] - \Tilde{\varrho}^{(3)} \epsilon^{\mu\nu\rho\sigma}k_\rho(v-v')_\sigma \\ 
    \Tilde{M}^{\mu\nu}_A&=\Tilde{\varrho}^{(1)}\mathcal{S}^\mu(v-v')^\nu+\Tilde{\varrho}^{(3)}\qty{ (v \cdot k)\mathcal{P}^{\mu\nu}+ (v' \cdot k)\mathcal{P}^{\prime\mu\nu}}\nn\\
    &-\Tilde{\varrho}^{(2)}\qty{(w+1)\mathcal{P}^{\mu\nu}-v'^\alpha \mathcal{P}^\mu_\alpha v^\nu }
    -\Tilde{\varrho}^{(2')}\qty{(w+1)\mathcal{P}^{\prime\mu\nu}-v^\alpha \mathcal{P}^{\prime\mu}_\alpha v'^\nu }.
\end{align}
\end{subequations}

\begin{subequations}
    \begin{align}
    &\Tilde{G}^\nu_{V,c}=0, &&\Tilde{G}^\nu_{V,b}=0,\\
    &\Tilde{G}^\nu_{A,c}=\frac{2 \tau_{1/2}}{v'\cdot k}(v-v')^\nu, &&\Tilde{G}^\nu_{A,b}=-\frac{2\tau_{1/2}}{v\cdot k}(v-v')^\nu,
\end{align}
\end{subequations}
\newline

\subsection{$\Bar{B}\to D_1^* l\Bar{\nu}\gamma$}
\begin{subequations}
    \begin{align}
    \Tilde{F}_V^\nu &= 2\tau_{1/2} [(w-1)\epsilon^{*\nu} -(v\cdot\epsilon^*) v'^\nu],\\ 
    \iu\Tilde{F}_A^\nu &=2\tau_{1/2} \epsilon^{\nu\alpha\rho\sigma}v_\alpha v'_\rho \epsilon^*_\sigma.
\end{align}
\end{subequations}

\begin{subequations}
    \begin{align}
    \Tilde{M}^{\mu\nu}_V&=\Tilde{\varrho}^{(1)}\mathcal{S}^\mu\qty[(w-1)\epsilon^{*\nu}-(v\cdot\epsilon)v'^\nu]\nn\\
    &-\Tilde{\varrho}^{(3)} \big\{(w-1)\epsilon^*_\alpha g^{\mu[\nu} k^{\alpha]} +v^{[\mu}v^{\prime\nu]} \qty(k \cdot \epsilon^*) - \qty( v' \cdot k)\qty(v \cdot \epsilon^*) g^{\mu\nu}\nn\\
    &\qquad \quad +\epsilon^{*\alpha}[(v'\cdot k)\mathcal{P}^{\prime\mu}_\alpha v^\nu+(v\cdot k)\mathcal{P}^\mu_\alpha v'^\nu ]+\epsilon^{*\alpha}[(v'\cdot k)\mathcal{P}^{\prime\nu}_\alpha v^\mu-(v\cdot k)\mathcal{P}^\nu_\alpha v'^\mu]\big\}\nn\\
    &-\Tilde{\varrho}^{(2)}\qty{(v\cdot\epsilon^*)\mathcal{P}^{\mu\nu}-\mathcal{P}^\mu_\alpha\epsilon^{*\alpha}(v-v')^\nu-\mathcal{P}^\mu_\alpha(v+v')^\alpha \epsilon^{*\nu}}\nn\\
    &-\Tilde{\varrho}^{(2')}\qty{(v\cdot\epsilon^*)\mathcal{P}^{\prime\mu\nu}-\mathcal{P}^{\prime\mu}_\alpha\epsilon^{*\alpha}(v-v')^\nu-\mathcal{P}^{\prime\mu}_\alpha(v+v')^\alpha \epsilon^{*\nu}},\\
    -i\Tilde{M}^{\mu\nu}_A&=-\Tilde{\varrho}^{(1)}\mathcal{S}^\mu\epsilon^{\nu\alpha\rho\sigma}v_\alpha v'_\rho \epsilon^*_\sigma + \epsilon^{\alpha\nu\rho\sigma}\qty(v'-v)_\rho\epsilon^*_\sigma \qty[\Tilde{\varrho}^{(2)}\mathcal{P}^\mu_\alpha +\Tilde{\varrho}^{(2')} \mathcal{P}^{\prime\mu}_\alpha]\nn \\
    &-\Tilde{\varrho}^{(3)}\big[\epsilon^{*\mu}\epsilon^{\nu\alpha\rho\sigma}v_\alpha v'_\rho k_\sigma+ \epsilon^{\mu\alpha\rho\sigma} k_\rho \epsilon^*_\sigma(v'^\nu v_\alpha+v^\nu v'_\alpha) -\epsilon^{\mu\nu\rho\sigma}[(w+1)k_\rho \epsilon^*_\sigma-(k\cdot \epsilon^*)v_\rho v'_\sigma]\big].
\end{align}
\end{subequations}

\begin{subequations}
    \begin{align}
    &\Tilde{G}^\nu_{V,c}=\frac{2\tau_{1/2}}{v'\cdot k}[(w-1)\epsilon^{*\nu}-(v\cdot \epsilon^*)v'^\nu],
    &&\Tilde{G}^\nu_{V,b}=-\frac{2\tau_{1/2}}{v\cdot k}[(w-1)\epsilon^{*\nu}-(v\cdot \epsilon^*)v'^\nu],\\ 
    &\iu\Tilde{G}^\nu_{A,c}=\frac{2\tau_{1/2}}{v'\cdot k}\epsilon^{\nu\alpha\rho\sigma}v_\alpha v'_\rho \epsilon^*_\sigma,
    &&\iu\Tilde{G}^\nu_{A,b}=-\frac{2\tau_{1/2}}{v\cdot k}\epsilon^{\nu\alpha\rho\sigma}v_\alpha v'_\rho \epsilon^*_\sigma.
\end{align}
\end{subequations}
\newline

\subsection{$\Bar{B}\to D_1 l\Bar{\nu}\gamma$}
\begin{subequations}
    \begin{align}
    \hat{F}_V^\nu &= -\frac{\tau_{3/2}}{\sqrt{2}}\qty{ \qty(v \cdot \epsilon^*) \qty[3 v^\nu-(w-2) v^{\prime\nu} ] +(w^2-1) \epsilon^{*\nu}}, \\
    -\iu\hat{F}_A^\nu &= \frac{\tau_{3/2}}{\sqrt{2}}\qty(w+1) \epsilon^{\nu\alpha\beta\rho}v_\alpha v'_\beta \epsilon^*_\rho.
\end{align}
\end{subequations}

\begin{subequations}
    \begin{align}
   \sqrt{6} \hat{M}^{\mu\nu}_V&= -\qty[\hat{\varrho}^{(1)} \mathcal{S}^\mu v_\sigma +\hat{\varrho}^{(4)}\mathcal{P}^{\mu}_{\sigma} + \hat{\varrho}^{(4')}\mathcal{P}^{\prime\mu}_{\sigma}]\qty{g^{\sigma[\nu}\epsilon^{*\alpha]}v_\alpha  +2 \epsilon^{*\sigma} \qty( v+ v')^\nu+ v^{\prime\sigma}\qty[w \epsilon^{*\nu} - v^{\prime\nu} \qty( v \cdot \epsilon^*)]} \nn \\
    &- \qty[v_\sigma \qty(\hat{\varrho}^{(2)} \mathcal{P}^\mu_\alpha + \hat{\varrho}^{(2')} \mathcal{P}^{\prime\mu}_\alpha) +\delta^\mu_{[\sigma} k_{\alpha]} \hat{\varrho}^{(5)} + k_\sigma \qty(\hat{\varrho}^{(6)} \mathcal{P}^\mu_\alpha + \hat{\varrho}^{(6')} \mathcal{P}^{\prime\mu}_\alpha)] \nn \\
    & \quad \times \qty{ (w-1) g^{\sigma [\nu} \epsilon^{*\alpha]} - v^\sigma v^{\prime[\nu}\epsilon^{*\alpha]} + 2 \epsilon^{*\sigma}\qty[(w-1)g^{\alpha\nu}- v^{[\alpha} v^{\prime\nu]}]-\qty(v \cdot \epsilon^*) v^{\prime[\alpha} g^{\nu]\sigma} - v^{\prime\sigma} v^{\prime[\alpha} \epsilon^{*\nu]}} \nn \\
    & +\qty[\hat{\varrho}^{(3)} v^\sigma + \hat{\varrho}^{(7)} k^\sigma] \Bigg\{ \epsilon^*_\alpha \qty[ \qty(v \cdot k) \mathcal{P}^{\mu}_\beta - \qty(v' \cdot k) \mathcal{P}^{\prime\mu}_\beta]\qty(g^{\nu[\alpha}\Pi^{\sigma]\beta}-2g^{\alpha\sigma}\Pi^{\beta\nu}) + v^{\prime\nu}\epsilon^*_\alpha g^{\mu[\sigma} k^{\alpha]} \nn \\
    & - \epsilon^*_\alpha g^{\mu[\sigma} k^{\nu} v^{\alpha]} +v^{\prime\sigma}\Big[\qty(k \cdot \epsilon^*)\qty(w\, g^{\mu\nu}- v^{\{\mu}v^{\prime\nu\}}) - \qty( v\cdot k)\mathcal{P}^\nu_\alpha v^{\prime[\mu}\epsilon^{*\alpha]} \nn \\
    & - \qty(v' \cdot k) \qty[ g^{\mu\nu} \qty( v\cdot \epsilon^*) - v^{\{\mu} \epsilon^{*\nu\}}] - v^\nu k_\alpha v^{\prime[\mu}\epsilon^{*\alpha]}\Big]\Bigg\}, \\
    -\iu \sqrt{6}\hat{M}^{\mu\nu}_A&=\qty[\hat{\varrho}^{(1)} \mathcal{S}^\mu v_\sigma +\hat{\varrho}^{(4)}\mathcal{P}^{\mu}_{\sigma} + \hat{\varrho}^{(4')}\mathcal{P}^{\prime\mu}_{\sigma}]\qty{\epsilon^{\nu\sigma\alpha\rho}\qty(v+v')_\alpha \epsilon^*_\rho+v^{\prime\sigma}\epsilon^{\nu\alpha\beta\rho}v_\alpha v'_\beta \epsilon^*_\rho} \nn \\
    & - \qty[v_\sigma \qty(\hat{\varrho}^{(2)} \mathcal{P}^\mu_\alpha + \hat{\varrho}^{(2')} \mathcal{P}^{\prime\mu}_\alpha) +\delta^\mu_{[\sigma} k_{\alpha]} \hat{\varrho}^{(5)} + k_\sigma \qty(\hat{\varrho}^{(6)} \mathcal{P}^\mu_\alpha + \hat{\varrho}^{(6')} \mathcal{P}^{\prime\mu}_\alpha)]  \nn \\
    & \quad \times \qty{\qty[\qty(v-v')^\sigma \epsilon^{\alpha\nu\rho\beta}+ 2 v^\alpha \epsilon^{\nu\sigma \rho \beta} + g^{\alpha\nu}\epsilon^{\sigma\lambda\rho\beta}v_\lambda]v'_\rho \epsilon^*_\beta  + 2\epsilon^{*\sigma} \epsilon^{\alpha\nu\rho\beta}v_\rho v'_\beta-(w-1)\epsilon^{\alpha\nu\sigma\beta}\epsilon^*_\beta } \nn \\
    & +\qty[\hat{\varrho}^{(3)} v_\sigma + \hat{\varrho}^{(7)} k_\sigma] \epsilon^*_\rho \, \Big\{-\frac{v \cdot k}{2}\qty[ \mathcal{P}^{[\mu}_\alpha \epsilon^{\nu]\alpha\sigma\rho} +  \mathcal{P}^{[\rho}_\alpha \epsilon^{\sigma]\alpha\mu\nu} + v^{\prime\sigma} \mathcal{P}^{[\mu\alpha}\epsilon^{\nu]\alpha\lambda\rho}v_\lambda] \nn \\
   & \qquad\qquad - \qty(v' \cdot k)\qty[\mathcal{P}^{\prime \{\mu}_\alpha \epsilon^{\nu\}\sigma\alpha\rho}+\mathcal{P}^{\prime \sigma}_\alpha \epsilon^{\mu\nu\alpha\rho} - v^{\prime\sigma}\mathcal{P}^{\prime \{\mu}_\alpha \epsilon^{\nu\}\alpha\lambda\rho}v_\lambda] \nn \\
   &\qquad\qquad + k_\alpha \qty(v+v')_\lambda  \qty[g^{\mu\nu}\epsilon^{\sigma\alpha\lambda\rho} -2 g^{\lambda\nu} \epsilon^{\mu\sigma\alpha\rho}+2g^{\sigma\rho}\epsilon^{\mu\nu\alpha\lambda} -g^{\sigma\alpha} \epsilon^{\mu\nu\lambda\rho}]  \nn \\
   &\qquad\qquad +k^\sigma \epsilon^{\mu\nu\lambda\rho}v_\lambda -k_\alpha v^{\prime\sigma} \qty[g^{\mu\nu}\epsilon^{\alpha\lambda\delta\rho} v_\lambda v'_\delta  - v^{[\mu}v^{\prime\lambda]}\epsilon^{\nu\lambda\alpha\rho}-  v^{[\nu}v^{\prime\lambda]}\epsilon^{\mu\lambda\alpha\rho}]\Big\}.
\end{align}
\end{subequations}

\begin{subequations}
    \begin{align}
    \hat{G}_{V,c}^{\nu}&=-\frac{\tau_{3/2}}{\sqrt{2}\qty(v' \cdot k)}\qty{ \qty(v \cdot \epsilon^*) \qty[3 v^\nu-(w-2) v^{\prime\nu} ] +(w^2-1) \epsilon^{*\nu}},\\
    \hat{G}_{V,b}^{\nu}&=\frac{\tau_{3/2}}{\sqrt{2}\qty(v \cdot k)}\qty{ \qty(v \cdot \epsilon^*) \qty[3 v^\nu-(w-2) v^{\prime\nu} ] +(w^2-1) \epsilon^{*\nu}},\\
    \iu\hat{G}_{A,c}^{\nu}&=-\frac{\tau_{3/2}}{\sqrt{2}(v' \cdot k)}\qty(w+1) \epsilon^{\nu\alpha\beta\rho}v_\alpha v'_\beta \epsilon^*_\rho, \\
    \iu\hat{G}_{A,b}^{\nu}&=\frac{\tau_{3/2}}{\sqrt{2}(v\cdot k)}\qty(w+1) \epsilon^{\nu\alpha\beta\rho}v_\alpha v'_\beta \epsilon^*_\rho.
\end{align}
\end{subequations}
\newline

\subsection{$\Bar{B}\to D_2^* l\Bar{\nu}\gamma$}
\begin{subequations}
    \begin{align}
    -\iu\hat{F}_V^\nu &= \sqrt{3}\tau_{3/2}\epsilon^*_{\mu\sigma}v^\mu\epsilon^{\nu\sigma\alpha\rho}v'_\alpha v_\rho,\\
    \hat{F}_A^\nu &= -\sqrt{3}\tau_{3/2}\epsilon^*_{\mu\sigma}v^\mu[(w+1)g^{\nu\sigma}-v'^\nu v^\sigma].
\end{align}
\end{subequations}

\begin{subequations}
    \begin{align}
    -\iu\hat{M}^{\mu\nu}_V&=\qty[\hat{\varrho}^{(1)} \mathcal{S}^\mu v^\sigma +\hat{\varrho}^{(4)}\mathcal{P}^{\mu\sigma} + \hat{\varrho}^{(4')}\mathcal{P}^{\prime\mu\sigma}]\epsilon^{*}_{\rho\sigma}\epsilon^{\nu\rho\alpha\beta}v'_\alpha v_\beta\nn\\ 
    &+ \qty[v_\sigma \qty(\hat{\varrho}^{(2)} \mathcal{P}^\mu_\alpha + \hat{\varrho}^{(2')} \mathcal{P}^{\prime\mu}_\alpha) +\delta^\mu_{[\sigma} k_{\alpha]} \hat{\varrho}^{(5)} + k_\sigma \qty(\hat{\varrho}^{(6)} \mathcal{P}^\mu_\alpha + \hat{\varrho}^{(6')} \mathcal{P}^{\prime\mu}_\alpha)] \epsilon^{*\sigma}_{\rho} \epsilon^{\nu\alpha\beta\rho} \qty(v -v')_\beta \nn\\  
    & +\qty[\hat{\varrho}^{(3)} v^\sigma + \hat{\varrho}^{(7)} k^\sigma] \Big\{\epsilon^*_{\sigma\alpha} \qty[\epsilon^{\alpha\rho\nu\beta}v'_\beta \qty(v \cdot k) \mathcal{P}^\mu_\rho + \epsilon^{\alpha\rho\nu\beta}v_\beta \qty(v' \cdot k) \mathcal{P}^{\prime\mu}_\rho - (w+1)\epsilon^{\alpha\mu\nu\rho}k_\rho]\nn \\
    & \qquad \qquad \qquad \qquad +\epsilon^*_{\sigma\nu} \epsilon^{\mu\alpha\beta\rho}k_\alpha v_\beta v'_\rho\Big\}, \\
    \hat{M}^{\mu\nu}_A&=\qty[\hat{\varrho}^{(1)} \mathcal{S}^\mu v^\sigma +\hat{\varrho}^{(4)}\mathcal{P}^{\mu\sigma} + \hat{\varrho}^{(4')}\mathcal{P}^{\prime\mu\sigma}]\epsilon^{*}_{\rho\sigma}\qty[v^\rho v'^\nu-(w+1)g^{\nu\rho}]\nn\\
    &+ \qty[v_\sigma \qty(\hat{\varrho}^{(2)} \mathcal{P}^\mu_\alpha + \hat{\varrho}^{(2')} \mathcal{P}^{\prime\mu}_\alpha) +\delta^\mu_{[\sigma} k_{\alpha]} \hat{\varrho}^{(5)} + k_\sigma \qty(\hat{\varrho}^{(6)} \mathcal{P}^\mu_\alpha + \hat{\varrho}^{(6')} \mathcal{P}^{\prime\mu}_\alpha)] \nn \\
    & \qquad  \times \qty[ \epsilon^{*\alpha}_{\sigma} \qty(v -v')^{\nu} - g^{\alpha\nu} \epsilon^*_{\sigma\rho} v^\rho + \epsilon^{*\nu}_\sigma \qty(v + v')^\alpha] \nn \\
    & +\qty[\hat{\varrho}^{(3)} v^\sigma + \hat{\varrho}^{(7)} k^\sigma] \Big[\epsilon^*_{\sigma \alpha} \qty(v' \cdot k)v^{\{\mu} \mathcal{P}^{\prime\nu\}\alpha} - \epsilon^*_{\sigma\rho} v^\rho \qty(v' \cdot k)\mathcal{P}^{\prime \mu\nu} \nn \\
    & \qquad \qquad + (w+1) g^{\mu[\nu} k^{\alpha]} \epsilon^*_{\sigma\alpha} + \qty(v \cdot k) \epsilon^{*[\mu}_\sigma v^{\prime \nu]}\Big].
\end{align}
\end{subequations}

\begin{subequations}
    \begin{align}
    &\iu\hat{G}_{V,c}^{\nu}=-\frac{\sqrt{3}\tau_{3/2}}{v'\cdot k}\epsilon^*_{\mu\sigma}v^\mu\epsilon^{\nu\sigma\alpha\rho}v'_\alpha  v_\rho,
    &&\iu\hat{G}_{V,b}^{\nu}=\frac{\sqrt{3}\tau_{3/2}}{v\cdot k}\epsilon^*_{\mu\sigma}v^\mu\epsilon^{\nu\sigma\alpha\rho}v'_\alpha  v_\rho,\\ 
    &\hat{G}_{A,c}^{\nu}=-\frac{\sqrt{3}\tau_{3/2}}{v'\cdot k}i\epsilon^*_{\mu\sigma}v^\mu[(w+1)g^{\nu\sigma}-v'^\nu v^\sigma],
    &&\hat{G}_{A,b}^{\nu}=\frac{\sqrt{3}\tau_{3/2}}{v\cdot k}i\epsilon^*_{\mu\sigma}v^\mu[(w+1)g^{\nu\sigma}-v'^\nu v^\sigma].
\end{align}
\end{subequations}

\section{Magnetic dipole moment operators}
\label{app:magnetic dipole moment operators}
In this appendix we essentially provide the Feynman rules for all the magnetic dipole operators present in \cref{eq:magnetic dipole contributions} defining the $\mathcal{O}(v)$ tensors introduced in \cref{subsec:sub-leading soft behavior}.
\begin{subequations}
    \begin{align}
        \mathcal{O}_h^\mu&=ie \mu_{h}^a\gamma^\mu\slashed{k},\\
        \mathcal{O}_k^\mu&=ie\mu_{k}^a\gamma^\mu\slashed{k},\\
        \mathcal{O}_{kh}^\mu&=\mathcal{O}_{hk}^\mu=ie\mu_{kh}^a\gamma^\mu\slashed{k},\\
        \mathcal{O}_{f}^{\mu\nu\rho}&=ie\mu_{f}^a g^{\nu\rho}\gamma^\mu\slashed{k},\\
        \mathcal{O}_{fh}^{\mu\nu}&=\mathcal{O}_{hf}^{\mu\nu}=\frac{ie\mu_{fh}^a}{2}\qty[(v\cdot k)g^{\mu\nu}- v^\mu k^\nu],\\
        \mathcal{O}_{fk}^{\mu\nu}&=\mathcal{O}_{kf}^{\mu\nu}=\frac{ie\mu_{fk}^a}{2}\qty[g^{\mu\nu}\slashed{k}- \gamma^\mu k^\nu],\\
        \mathcal{O}_{r}^{\mu\nu\rho}&=\frac{ie\mu_r^a}{2}\qty(g^{\mu\nu}k^{\rho}- g^{\mu\rho}k^\nu),\\
        \mathcal{O}_{rt}^{\mu\nu}&=\mathcal{O}_{tr}^{\mu\nu}=\frac{ie\mu_{rt}^a}{2}\qty[(v\cdot k)g^{\mu\nu}- v^\mu k^\nu].
    \end{align}
\end{subequations}
$a$, as in the main text, is a $SU(3)_V$ index.

\section{Additional Isgur-Wise functions}
\label{app:additional Isgur-Wise functions}
In order to compute the contribution of some of the higher order operators in \hhchpt presented in \cref{eq:magnetic dipole contributions}, we would be required to compute the Isgur-Wise functions for transitions among excited states. Even though we are not interested in those suppressed contributions in our analysis, for completeness we provide  the Isgur-Wise functions for the following additional processes: $\Lambda_{b1}^{(*)}\to \Lambda_{c1}^{(*)}l\bar{\nu}$ and $\bar{B}^{**}\to D^{**} l\bar{\nu}$.

For the $\Lambda_{b1}^{(*)}\to \Lambda_{c1}^{(*)}$ transition, we have
\begin{equation}
    F'_{\Gamma}=\qty[\sigma_1'(w) g^{\alpha\beta} + \sigma_2'(w) v_\alpha v'_\beta]\mel{\Lambda_{c1}^{(*)}(v')}{ \overline{\mathcal{R}}_{v'}^{(c)\alpha}\Gamma \mathcal{R}_{v}^{(b)\beta}}{\Lambda_{b1}^{(*)}(v)},
\end{equation}
with two new LO Isgur-Wise functions $\sigma_{1,2}'(w)$, with $\sigma_1'(1)=1$ and $\sigma_2'(1)$ unconstrained.

The transition $\bar{B}^{**}\to D^{**}l\bar{\nu}$ includes four different processes, out of which three are independent. As far as $\bar{B}^{1/2^+}\to D^{1/2^+} l\bar{\nu}$
\begin{equation}
    \Tilde{F}'_{\Gamma}=-\Tilde{\xi}'(w)\mel{ D^{1/2^+}(v')}{\Tr{ \overline{\mathcal{K}}_{v'}^{(c)}\Gamma \mathcal{K}_{v}^{(b)}}}{\bar{B}^{1/2^+}(v)},
\end{equation}
so that there is a new IW function $\Tilde{\xi}'(w)$, that at zero recoil is normalized to unity. 

Looking at $\bar{B}^{3/2^+}\to D^{3/2^+} l\bar{\nu}$ one has two new LO IW functions\footnote{Also in this case the term proportional to $\epsilon^{\sigma\rho\alpha\beta}v_\alpha v'_\beta \gamma^5$ turns out to be redundant.},
\begin{equation}
    \hat{F}'_{\Gamma}=-\qty[ \hat{\xi}_1'(w) g^{\alpha\beta}+ \hat{\xi}_2'(w) v_\alpha v'_\beta]\mel{ D^{3/2^+}(v')}{\Tr{ \overline{\mathcal{F}}_{v'}^{(c)\alpha}\Gamma \mathcal{F}_{v}^{(b)\beta}}}{\bar{B}^{3/2^+}(v)},
\end{equation}
with $\hat{\xi}_1'(1)=1$ and $\hat{\xi}_2'(1)$ unconstrained.

Finally, for $\bar{B}^{3/2^+}\to D^{1/2^+} l\bar{\nu}$ or $\bar{B}^{1/2^+}\to D^{3/2^+} l\bar{\nu}$, we have again a single Isgur-Wise function $\hat{\tau}'(w)$,
\begin{equation}
    \hat{F}''_\Gamma=-\sqrt{3}\hat{\tau}'(w)\mel{D^{3/2^+}(v')}{\Tr{v_\mu\overline{\mathcal{F}}_{v'}^{(c)\mu}\Gamma \mathcal{K}_v^{(b)}}}{\Bar{B}^{1/2^+}(v)}.
\end{equation}
Since at zero recoil $\hat{F}''^\nu_\Gamma$ vanishes, $\hat{\tau}'(1)$ is unconstrained as expected.

\bibliographystyle{JHEP}
\bibliography{biblio.bib}
\end{document}